 \documentclass[final,5p,times,twocolumn,authoryear]{elsarticle}

\usepackage{amssymb}
\usepackage{lipsum}
\usepackage{amsmath,amsthm,mathtools}
\usepackage{xcolor}
\journal{Physics Letters A}

\begin{document}

\begin{frontmatter}

%% Title, authors and addresses

%% use the tnoteref command within \title for footnotes;
%% use the tnotetext command for theassociated footnote;
%% use the fnref command within \author or \affiliation for footnotes;
%% use the fntext command for theassociated footnote;
%% use the corref command within \author for corresponding author footnotes;
%% use the cortext command for theassociated footnote;
%% use the ead command for the email address,
%% and the form \ead[url] for the home page:
%% \title{Title\tnoteref{label1}}
%% \tnotetext[label1]{}
%% \author{Name\corref{cor1}\fnref{label2}}
%% \ead{email address}
%% \ead[url]{home page}
%% \fntext[label2]{}
%% \cortext[cor1]{}
%% \affiliation{organization={},
%%            addressline={}, 
%%            city={},
%%            postcode={}, 
%%            state={},
%%            country={}}
%% \fntext[label3]{}

\title{Systematic 2.5 D resistive MHD simulations with ambipolar diffusion and Hall effect for fast magnetic reconnection}

%% use optional labels to link authors explicitly to addresses:
%% \author[label1,label2]{}
%% \affiliation[label1]{organization={},
%%             addressline={},
%%             city={},
%%             postcode={},
%%             state={},
%%             country={}}
%%
%% \affiliation[label2]{organization={},
%%             addressline={},
%%             city={},
%%             postcode={},
%%             state={},
%%             country={}}

\author[first]{Gabriela Landinez}
\affiliation[first]{organization={Universidad Industrial de Santander},%Department and Organization
            addressline={A. A. 678}, 
            city={Bucaramanga},
            postcode={680002}, 
            state={Santander},
            country={Colombia}}

\author[first]{Fabio D. Lora-Clavijo}
%\affiliation[first]

\begin{abstract}
In this work, we explore the possibility of the Hall effect and ambipolar diffusion as a mechanism for fast reconnection. The reconnected flux of our resistive and resistive+Hall simulations replicates the GEM results. Furthermore, we investigate, for the first time, the effect of ambipolar diffusion in the GEM. The reconnected flux of the resistive+ambipolar and resistive+Hall+ambipolar simulations showed increases of up to 75\% and 143\%, respectively, compared to the resistive and resistive+Hall simulations, showing that ambipolar diffusion contributes significantly to the reconnected flux. Our second scenario has a magnetic Harris field without perturbations but with an out-of-plane component, known as a guide field. We found that the reconnection rate increased faster with ambipolar diffusion, reaching values close to 0.1 for the resistive+Hall+ambipolar simulation followed by the resistive+Hall. These two simulations achieved the highest kinetic energy, implying more efficient energy conversion during reconnection.
\end{abstract}

%%Graphical abstract
%\begin{graphicalabstract}
%\includegraphics{grabs}
%\end{graphicalabstract}

%%Research highlights
%\begin{highlights}
%\item Research highlight 1
%\item Research highlight 2
%\end{highlights}

\begin{keyword}
%% keywords here, in the form: keyword \sep keyword, up to a maximum of 6 keywords
magnetic reconnection \sep Hall MHD \sep ambipolar diffusion \sep numerical simulations \sep numerical methods

%% PACS codes here, in the form: \PACS code \sep code

%% MSC codes here, in the form: \MSC code \sep code
%% or \MSC[2008] code \sep code (2000 is the default)

\end{keyword}

\end{frontmatter}

%\tableofcontents

%% \linenumbers

%% main text

\section{Introduction}
\label{introduction}

Magnetic reconnection is a topological rearrangement of the magnetic field that converts magnetic energy into plasma energy \citep{Zweibel2009}. This process is usually described in terms of a change in the connectivity of the field lines, also known as the magnetic field topology, which allows the release of magnetic energy stored in the system. The Ohmic resistivity is responsible for this release, causing the plasma frozen-in-flux condition to be violated in the diffusion zone, where breaking and reconnection of magnetic field lines occur \citep{Priest2000, Bittencourt2013}. Magnetic reconnection takes place in different plasmas, spanning from laboratory to astrophysical environments, is responsible for the different mechanisms of particle acceleration, the main driver of space weather \citep{Priest2000, Birn2012} and astrophysical phenomena, such as geomagnetic storms. To understand the energy conversion associated with these phenomena, it is essential to know the rate at which magnetic reconnection occurs \citep{comisso2016value}. For example, the amplitude of geomagnetic disturbances can be controlled by the rate of magnetic reconnection \citep{Nakamura2018}, which provides information about how fast the process occurs, how much magnetic flux is reconnected, and how much magnetic energy becomes kinetic energy that accelerates the plasma \citep{Hesse2020}. For these reasons, the magnetic reconnection rate has been studied observationally, experimentally, theoretically, and numerically. Nowadays, it is known from numerical simulations and satellite observations that the normalized reconnection rate is about 0.1 in many systems \citep{Cassak2017, Hesse2020}, a much larger value than that predicted by the classical model of Sweet \citep{Sweet1958a, Sweet1958b} and Parker \citep{Parker1957, Parker1963}. However, despite years of effort and first-principles theories \citep{liu2022, goodbred2022}, a complete theoretical understanding has not been achieved. The above suggests that those classical models are insufficient to capture all the relevant processes during reconnection, probably because the physical mechanisms for fast reconnection are challenging to study through theoretical means \citep{malyshkin2011onset}.

The key to explaining a value of 0.1 may lie in studying the particle decoupling caused by the Hall effect, or in other physical phenomena beyond the single-fluid magnetohydrodynamic (MHD) approximation, like ambipolar diffusion. The GEM reconnection challenge \citep{GEM2001} looked into that and verified that all their models with the Hall effect had similar reconnection rates, but later other simulations \citep{Karimabadi2004, Bessho2005, Bessho2007} showed that reconnection also has a similar rate without this effect. However, most of these works are not concerned with partially ionized plasmas. According to \cite{malyshkin2011onset}, in a weakly ionized plasma with significant interaction between ions and neutrals, fast reconnection is triggered by the Hall effect at considerably lower Lundquist number values compared to scenarios with fully ionized plasma. These conditions for fast reconnection are satisfied in molecular clouds, where another effect could play a key role: the ambipolar diffusion, which is extremely important in astrophysics and contributes significantly to events of interstellar medium heating \citep{Falgarone2015, Zweibel2015}. So far, simulations have shown that ambipolar diffusion causes thinning of the current sheet, favouring rapid reconnection rates \citep{Ethan1999, Heitsch2003, Ni2015}. However, the direct effect of ambipolar diffusion on the reconnection rate has yet to be well studied in the Earth's magnetotail. While the literature on the value $0.1$ for partially ionized plasmas is limited, numerical kinetic simulations suggest that magnetic reconnection rates can approach this value \citep{jara2019kinetic}. Although research on magnetic reconnection in partially ionized plasmas is less extensive, a multitude of astrophysical environments, including the lower solar atmosphere, comet tails, protoplanetary nebulae, and disks around young stellar objects, are filled with partially ionized plasmas. This motivates further investigation into the underlying mechanisms driving fast reconnection and the conditions under which it can occur in these environments.

Even in the presence of neutrals, all plasma components can be assumed to be strongly coupled by collisions, and the single fluid remains adequate. In many cases, however, collisions between particles are not sufficient to fully couple all species in a plasma. For example, in diffusion regions where magnetic reconnection occurs, the single-fluid approximation is no longer valid: ions are decoupled from the magnetic field and no longer move with electrons. Similarly, there are regions where the neutral gas in a plasma separates from the charged particles. This gives rise to relative velocities that manifest themselves as non-ideal processes, such as the Hall effect and ambipolar diffusion \citep{Khomenko2012}. Fortunately, even when the plasma is modelled as a single fluid, it is possible to adopt an approach to deal with the effects produced by the interaction between its species. This approach, instead of including more fluids in the system, relies on the use of a generalized Ohm's law, whose additional terms contain the necessary information to describe a plasma where the different species are not strongly coupled.

With the above motivation, we address the problem of magnetic reconnection rate by performing a systematic comparison of simulations. We take a single-fluid approach with terms associated with the Hall effect and ambipolar diffusion in the resistive MHD equations. We do this to determine the effect of both phenomena on the behaviour of magnetic reconnection, in particular on the amount of reconnected flux and the reconnection rate. All numerical simulations are obtained by implementing and modifying subroutines in the MAGNUS code \citep{Navarro2017}. The paper is structured as follows. In section \ref{sec:equations} we present the resistive MHD equations with the Hall effect and ambipolar diffusion, which are included in the equations through Ohm's law. In section \ref{sec:magnus} we give a brief description of MAGNUS and the inclusion of both effects in the code. Then, in section \ref{sec:test} we perform the GEM benchmark test for the Hall effect. To our knowledge, no one has used the GEM to study the effect of ambipolar diffusion on the reconnected flux, so in section \ref{sec:ambipolarGEM} we present simulations with ambipolar diffusion using the GEM model too. In section \ref{sec:cs} we present the results for another simple current sheet model with guide field, in which we measure reconnection rates, magnetic and kinetic energies, and fluxes related to the energy transport equation. Finally, in section \ref{sec:conclusions} we summarize the main conclusions.

\section{MHD equations for a conducting fluid with Hall and ambipolar terms}
\label{sec:equations}

As mentioned above, we take a single-fluid approach with Hall and ambipolar diffusion terms in the resistive MHD equations. The MHD model describes low-frequency interactions between conducting fluids and electromagnetic fields, i.e. motions where $u^{2}/c^{2}<<1$, where $u$ is the characteristic fluid velocity and $c$ is the speed of light \citep{Schnack2009}.

\subsection{Ohm's law}

Ohm's law couples the dynamics of the fluid and the electromagnetic fields, taking into account non-ideal MHD processes beyond the ohmic resistance. For a reference frame in which the fluid moves with velocity $\mathbf{u}$, Ohm's law reads as follows \citep{Bittencourt2013,Ballester2018}
\begin{equation}
    \mathbf{E} + \mathbf{u} \times \mathbf{B} = \eta \mathbf{j} + \eta_{H} (\mathbf{j} \times \mathbf{B}) - \eta_{A} \left[ (\mathbf{j} \times \mathbf{B}) \times \mathbf{B} \right],
\label{eq.Ohm}
\end{equation}
where $\eta$ is the ohmic resistivity, $\eta_{H}$ is the Hall coefficient, and $\eta_{A}$ is the ambipolar diffusion coefficient. The coefficients are defined as 
\begin{equation}
    \eta_{H} = \frac{1}{ne}, \qquad
    \eta_{A} = K_{A} \cdot \frac{1}{\rho^{2}\sqrt{T}},
\end{equation}
where $n$ is the number density, $e$ is the electron charge, $\rho$ is the plasma density, and $T$ is the temperature. $K_{A}$ is a parameter that controls the effect of the ambipolar term in our simulations and depends on the collision frequencies between particles and the physical conditions of the plasma \citep{Vigano2019}. For example, $K_{A}$ is inversely related to the degree of ionization, so for a fully ionized plasma, $K_{A}$ becomes zero and we do not have ambipolar diffusion. Ohm's law is often derived from the electron momentum equation under simplifying assumptions, such as neglecting electron inertia, gravity acting on electrons, and assuming that currents vary much more slowly in time than hydrodynamic processes \citep{2011A&A...529A..82Z}. Additionally, strong couplings, which are sometimes broken in partially ionized plasmas \citep{brandenburg1994formation, 1995ApJ...448..734B}, are often assumed. For a detailed discussion of Ohm's law in multi-component partially ionized plasmas, see \citep{khomenko2014fluid}. In this research, we focus specifically on analyzing the impact of Hall terms and ambipolar diffusion on the plasma dynamics associated with magnetic reconnection phenomena.

\subsubsection{Hall effect}

The Hall effect refers to the appearance of an electric field due to charge separation in a conductor. In plasma, the separation occurs as a consequence of decoupling between ions and electrons that takes place in the diffusion region, where a multiscale structure is developed based on the characteristic lengths of each species.

The Hall term arises from the separation of electron and ion motions in the plasma. In the generalized Ohm's law, this term captures the dynamics of the magnetic field at scales smaller than the ion inertial length. Including the Hall term allows for a faster reconnection rate than predicted by classical MHD. This is essential for explaining the rapid energy release observed in many astrophysical and laboratory plasmas. The Hall term leads to the formation of quadrupolar magnetic fields around the reconnection site. These structures are key observational signatures in both simulations and experiments.

\begin{figure}[h]
\begin{center}
\noindent\includegraphics[width=0.48\textwidth]{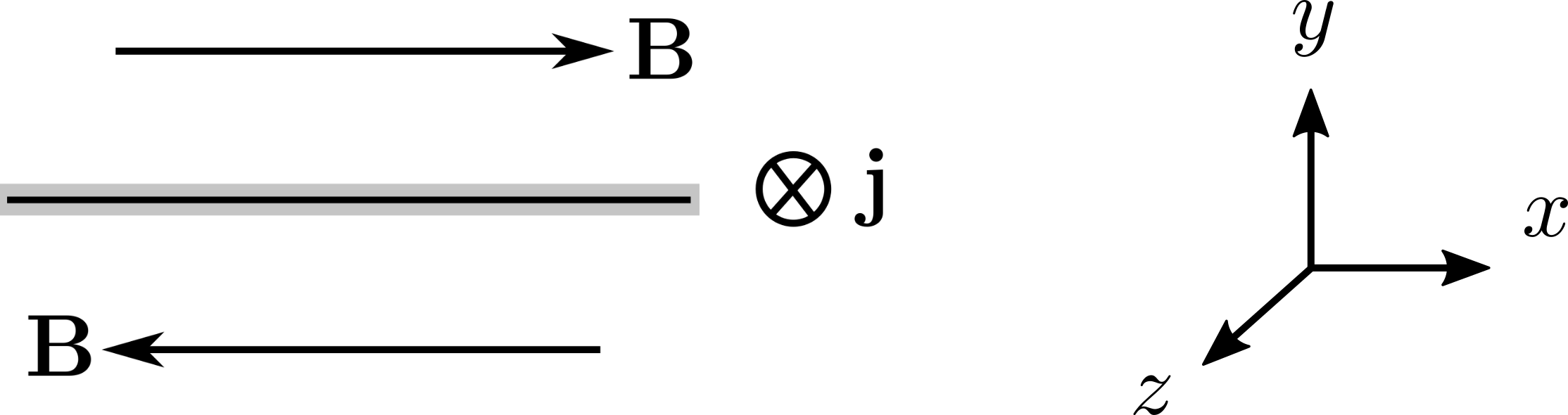}
\caption{Simple diagram of a current sheet. The magnetic field is located in the $xy$ plane and the current sheet in the $xz$ plane is oriented towards the $-z$ axis.}
\label{fig0}
\end{center}
\end{figure}

%To analyse the additional electric field that appears in Ohm's law (\ref{eq.Ohm}) due to the Hall term, we will use a simple and specific magnetic field configuration.
Given an initial antiparallel magnetic field as the one in Figure \ref{fig0}, the term $\mathbf{j} \times \mathbf{B}$ implies the appearance of electric fields in the $xy$ plane, where reconnection occurs.
To understand the dynamics of ions and electrons influenced by these electric fields and the appearance of the quadrupole field, a characteristic signature of Hall reconnection, the reader may refer to the physical picture made by \cite{Uzdensky2006}. Their description provides a clear example of how currents induced by the Hall effect may affect the structure of the initial current sheet as well as the magnetic field and its evolution \citep{Morales2020}.

%\textcolor{red}{To explain the generation of currents due to that electric field, we consider a simple physical picture made by \cite{Uzdensky2006} (for clarity when reading the following lines, we kindly request the reader to refer to Figure 1 of their paper)}. When reconnection is about to occur and the incoming plasma flux tube penetrates the diffusion region, the magnetic field decreases, causing an expansion in the central part of the flux tube and hence a reduction in the electron density, while ions are not magnetized and their density remains unaffected. However, because of the need to maintain charge neutrality, an electric field which pulls the electrons along the field lines appears. \textcolor{red}{As a result, there is a strong inflow of electrons along the magnetic field, moving from the outer flux tube toward its central region. Then, when reconnection occurs, and a reconnected field line moves away, the volume of its central part decreases and so the electrons flow rapidly outward along the field. Of course, there is an electric current associated with the flow of electrons. Moreover, by Ampere’s law, this current generates a quadrupole toroidal magnetic field, a characteristic signature of the Hall regime of reconnection}. These additional currents caused by the Hall effect may affect the structure of the initial current sheet as well as the magnetic field and its evolution \citep{Morales2020}.

\subsubsection{Ambipolar diffusion}

Ambipolar diffusion affects the transport of magnetic flux and can lead to the decoupling of magnetic fields from the bulk plasma. This decoupling is crucial for understanding the reconnection process in environments where neutrals are present.

Ambipolar diffusion also produces the appearance of electric fields, this time due to decoupling between neutrals and charged particles \citep{Zweibel2015}. The ambipolar term $(\mathbf{j} \times \mathbf{B}) \times \mathbf{B}$ generates electric fields in the $xz$ plane of Figure \ref{fig0}, as a consequence it affects the perpendicular currents to the reconnection plane. According to \cite{Morales2020}, this implies that ambipolar diffusion cannot produce magnetic reconnection. However, it can relax the magnetic field configuration by reducing the magnetic energy, which is then converted to heat by a dissipation process.

Ambipolar diffusion influences the reconnection rates by modifying the resistivity in the plasma. Depending on the local plasma conditions, this can either enhance or suppress the reconnection rate. In astrophysical contexts, such as star formation and the evolution of molecular clouds. Understanding its impact is necessary for realistic modelling of these processes.

Although the Hall effect generates currents in the plane where reconnection occurs, it does not contribute directly to reconnection because it continues to freeze the magnetic field in the electron flow \citep{Priest2000}. In fact, both effects, unlike ohmic resistance, preserve the magnetic field topology and do not trigger reconnection. However, they can promote its occurrence and affect its rate with effects such as current sheet thinning and plasmoid formation \citep{Nobrega2020, Zweibel2015}. Including both terms in the generalized Ohm's law lead to better predictions of reconnection rates, energy release, and the overall behaviour of plasma in both astrophysical and laboratory settings.

Finally, it is important to clarify that the appearance of electric fields in the plane where reconnection occurs, or in its perpendicular plane, will change depending on the configuration adopted, e.g. if a guide field is used, the $z$ component of the magnetic field must be considered. Figure \ref{fig0} is only used to illustrate a simple and specific configuration.
%Moreover, if there is a strong guiding field the assumption of ions being unmagnetized may not be valid \citep{Uzdensky2006}. 

\subsection{Transport equations}

When the plasma, which may contain ions, electrons, and neutrals, is considered as a single conducting fluid, the transport equations for each species are replaced by transport equations for the whole plasma, meaning that variables such as density and velocity for each species are replaced by average values \citep{Bittencourt2013}. In conservative form, the macroscopic transport equations for a conducting fluid are
\begin{gather}
\frac{\partial \rho}{\partial t} + \nabla \cdot (\rho \mathbf{u}) = 0,
\label{eq.mass} \\ \nonumber \\
\frac{\partial \rho \mathbf{u}}{\partial t} + \nabla \cdot \left( \rho \mathbf{u}\mathbf{u} - \frac{1}{\mu_{0}} \mathbf{B}\mathbf{B} + P \mathbf{I} \right) = 0,
\label{eq.momentum} \\ \nonumber \\
\frac{\partial E}{\partial t} + \nabla \cdot \left\lbrace (E + P) \mathbf{u} - \frac{1}{\mu_{0}} (\mathbf{B} \cdot \mathbf{u}) \mathbf{B} + \frac{1}{\mu_{0}} \left[ B^{2}\mathbf{u}_{H} - (\mathbf{B} \cdot \mathbf{u}_{H})\mathbf{B} \right] \right. \nonumber \\ \left. + \frac{1}{\mu_{0}} ( B^{2}\mathbf{u}_{A})  \right\rbrace = - \nabla \cdot \left( \frac{\eta}{\mu_{0}} \mathbf{j} \times \mathbf{B} \right), 
\label{eq.energy}
\end{gather}
where $\mathbf{u}_{H}$ and $\mathbf{u}_{A}$ are velocities associated with the Hall effect and the ambipolar diffusion, defined as
\begin{equation}
    \mathbf{u}_{H}=-\eta_{H}\mathbf{j}, \qquad
    \mathbf{u}_{A}=\eta_{A}(\mathbf{j} \times \mathbf{B}).
    \label{eq.coeff}
\end{equation}
The total pressure $P$ is defined as 
\begin{equation}
    P = p + \frac{B^{2}}{2\mu_{0}},   
\end{equation} 
the sum of fluid pressure $p$ and the magnetic pressure $B^{2}/2\mu_{0}$. The total energy density $E$ is given by the sum of the kinetic, magnetic and internal energy densities, that is
\begin{equation}
    E = \frac{1}{2} \rho u^{2} + \frac{1}{2\mu_{0}} B^{2} + \rho e,
    \label{eq.totalE}
\end{equation}
here $e$ is the system's internal energy and is related to pressure by the equation of state
\begin{equation}
    p = (\gamma - 1)\rho e,
\end{equation}
where $\gamma$ is the adiabatic index. The equations (\ref{eq.mass}), (\ref{eq.momentum}) and (\ref{eq.energy}) correspond to the transport equations for mass, momentum and energy. To obtain them, no gravitational force is assumed, viscosity and heat flow are neglected.

\subsection{Evolution equations for the electromagnetic fields}

Evolution equations for $\mathbf{E}$ and $\mathbf{B}$ are given by Faraday's and Ampère-Maxwell's laws, 
\begin{gather}
    \nabla \times \mathbf{E} = -\frac{\partial \mathbf{B}}{\partial t},
    \label{eq.Faraday} \\ \nonumber \\
    \nabla \times \mathbf{B} = \mu_{0} \mathbf{j},
    \label{eq.Ampere}
\end{gather}
where the displacement current term in equation (\ref{eq.Ampere}) is neglected due to the low-speed approximation. Also, at all times it must be guaranteed that
\begin{equation}
    \nabla \cdot \mathbf{B} = 0,
    \label{eq.freediv}
\end{equation}
which also guarantees the absence of magnetic monopoles. Finally, Faraday's law can be rewritten using Ohm's law to obtain the induction equation
\begin{equation}
    \frac{\partial \mathbf{B}}{\partial t} + \nabla \cdot (\mathbf{u}\mathbf{B} - \mathbf{B}\mathbf{u}) + \nabla \cdot (\mathbf{u}_{H}\mathbf{B} - \mathbf{B}\mathbf{u}_{H}) + \nabla \cdot (\mathbf{u}_{A}\mathbf{B} - \mathbf{B}\mathbf{u}_{A}) = - \nabla \times \eta \mathbf{j}.
    \label{eq.induction}
\end{equation}
With equations (\ref{eq.freediv}) and (\ref{eq.induction}), the system of equations (\ref{eq.mass})-(\ref{eq.energy}) is closed and we can proceed with the numerical methods used to solve it.

\section{Numerical methods: MAGNUS code}
\label{sec:magnus}

MAGNUS \citep{Navarro2017} is the Newtonian version of the relativistic code CAFE \citep{Lora2015}. It solves the resistive MHD equations and has been used mainly to study wave propagation in the solar atmosphere. Some of the scenarios that have been simulated with MAGNUS are: the emerging plasma blob in a solar coronal hole \citep{Navarro2019}, the thermal conduction effects on the formation of tadpole-like jets in the solar chromosphere \citep{Navarro2021}, and the propagation of torsional Alfvén waves in a stratified solar atmosphere \citep{Wandurraga2021}. More recently, an adaptation of MAGNUS has been used to model the propagation of PS-V seismic waves \citep{Landinez2021}.

The code solves the system of equations in a uniform grid using the method of lines, whose main idea is to replace the spatial derivatives by algebraic approximations \citep{Schiesser2009}. Once this is done, any integration method can be used to compute a solution. In MAGNUS, the algebraic approximation of the right-hand side of the equations is done using the finite volume method in combination with the HLLE approximate Riemann solver, which uses different slope limiters. The time integration can be done with different integrators of the Runge-Kutta family implemented in MAGNUS. In particular, we use the MC beta limiter with $\beta=1.2$ and a TVD second-order Runge-Kutta in all simulations.

The finite volume discretization of the MHD equations, including Hall and ambipolar terms, modifies the induction equation and the total energy density equation. We found that it is not possible to consider the Hall term as an additional source term, therefore it is necessary to add new terms to the numerical fluxes.

\subsection{Hall and ambipolar corrections to the fluxes}

With the velocities $\mathbf{u}_{H}$ and $\mathbf{u}_{A}$, the energy and induction equations can be rewritten in coordinates as
\begin{gather}
    \frac{\partial E}{\partial t}  =  -\frac{\partial}{\partial x_{j}} \left[ (E + P) u_{j} - \frac{1}{\mu_{0}} (\mathbf{B} \cdot \mathbf{u}) B_{j} \right] \nonumber  \\ - \frac{\partial}{\partial x_{j}} \left[ B^{2}u^{H}_{j} - (\mathbf{B} \cdot \mathbf{u}^{H})B_{j} \right] \nonumber \\  - \frac{\partial}{\partial x_{j}} \left( B^{2} u^{A}_{j} \right) - \left[ \nabla \cdot \left( \frac{\eta}{\mu_{0}} \mathbf{j} \times \mathbf{B} \right) \right]_{j},
\label{flux.e}
\end{gather}

\begin{gather}
    \frac{\partial B_{i}}{\partial t}  =  - \frac{\partial}{\partial x_{j}} (u_{j}B_{i} - B_{j}u_{i}) - \frac{\partial}{\partial x_{j}} (u^{H}_{j}B_{i} - B_{j}u^{H}_{i}) \nonumber \\  - \frac{\partial}{\partial x_{j}} (u^{A}_{j}B_{i} - B_{j}u^{A}_{i}) - [\nabla \times \eta \mathbf{j}]_{i}.
    \label{flux.i}
\end{gather}
The magnetic permeability $\mu_{0}$ does not appear in the above equations because MAGNUS solves the dimensionless system of equations.

The first three terms on the right-hand side of both equations are fluxes, while the others are source terms. Since MAGNUS already solves the resistive MHD equations, only the second and third terms in parentheses need to be added to the fluxes of each equation. With the finite volume method, the equations are discretized as
 \begin{gather}
     \frac{d \mathbf{U}_{(i, j, k)}}{d t}  =  -\frac{\mathbf{F}_{(i+1 / 2, j, k)}^x-\mathbf{F}_{(i-1 / 2, j, k)}^x}{\Delta x} \nonumber -\frac{\mathbf{F}_{(i, j+1 / 2, k)}^y-\mathbf{F}_{(i, j+1 / 2, k)}^y}{\Delta y} \nonumber \\ -\frac{\mathbf{F}_{(i, j, k+1 / 2)}^z-\mathbf{F}_{(i, j, k-1 / 2)}^z}{\Delta z}+\mathbf{S}_{(i, j, k)},
 \end{gather}
where $\mathbf{U}$ is the state vector, $\mathbf{F}^{i}$ the vector of fluxes along each direction\footnote{By using $1/2$ in one of the subindexes of the triplet $(i,j,k)$, we denote a face between cells in a given direction. For example, The subindex $(i+1/2,j,k)$ represents the face between cells in $i$ direction, that is, the face between $(i,j,k)$ and $(i+1,j,k)$.}, and $\mathbf{S}$ the vector of source terms. In MAGNUS, each vector of fluxes is calculated as
\begin{equation}
    \mathbf{F}_{i\pm1/2,j,k} = \mathbf{F}^{HLLE}_{i\pm1/2,j,k} + \mathbf{F}^{H}_{i\pm1/2,j,k} + \mathbf{F}^{A}_{i\pm1/2,j,k},
    \label{eq.flux}
\end{equation}
where $F^{HLLE}$ is the numerical flux for the MHD part, calculated using the HLLE High-Resolution Shock Capturing method. $F^{H}$ and $F^{A}$ are the fluxes due to Hall and ambipolar terms, both computed in different subroutines and then added to the other fluxes computed with the HLLE scheme along each spatial direction. Equation (\ref{eq.flux}) is analogous for $F_{i,j\pm1/2,k}$, $F_{i,j,k\pm1/2}$.

To compute the Hall and ambipolar velocities, which appear in (\ref{flux.e}) and (\ref{flux.i}), we need the magnetic field and the current density evaluated at the intercell. Since the current density is given by $\mathbf{j}=(\nabla \times \mathbf{B})/\mu_{0}$, we need the spatial derivatives of the magnetic field. For the latter, some authors propose schemes where the normal derivatives are computed differently from the tangential derivatives (see \cite{Toth2008} and \cite{Strumik2017}). However, we compute the derivatives of the magnetic field along each direction using a second-order finite-difference scheme. At each of the nodes, the derivatives are numerically approximated using forward, backwards, and central finite differences. Once this is done, the Hall and ambipolar velocities can be obtained at each point in the numerical grid. We then use the averaged variables
\begin{eqnarray}
     (u^{H})_{i+1/2,j,k} = \frac{(u^{H})_{i,j,k} + (u^{H})_{i+1,j,k}}{2}, \\
     (u^{A})_{i+1/2,j,k} = \frac{(u^{A})_{i,j,k} + (u^{A})_{i+1,j,k}}{2},
 \end{eqnarray}
to calculate the velocities at the intercell. The same is done for $(u^{H})_{i,j+1/2,k}$, $(u^{H})_{i,j,k+1/2}$, $(u^{A})_{i,j+1/2,k}$ y $(u^{A})_{i,j,k+1/2}$ and also for each component of the magnetic field in equations (\ref{flux.e}) and (\ref{flux.i}).

\subsection{Time-step modification}

Following the Courant-Friedrichs-Levy condition, the time step in MAGNUS is chosen as
 \begin{equation}
    \Delta t = C_{CFL} \times \min \left(\frac{\Delta x}{\left|\lambda_{i j k}^{n, x}\right|}, \frac{\Delta y}{\left|\lambda_{i j k}^{n, y}\right|}, \frac{\Delta z}{\left|\lambda_{i j k}^{n, z}\right|}\right),
\label{dt.resistive}
 \end{equation}
where the Courant-Friedrichs-Levy parameter $C_{CFL}$ must be less than or equal to 1, and $\lambda_{i j k}^{n, d}$ is the velocity of the fastest wave propagating in some $d$ direction at a time level $n$ \citep{Navarro2017}, so that the time step adapts at each level.

In the presence of Hall and ambipolar terms, some modifications are required. To do this, we performed a dimensional analysis of each of these terms in the induction equation. For the Hall term, we have
 \begin{equation}
    \frac{B_{o}}{t_{o}} = \eta_{H} \frac{j_{o}B_{o}}{x_{o}} = \eta_{H} \frac{B_{o}^{2}}{x_{o}^{2}},
 \end{equation}
therefore
 \begin{equation}
    t_{o} = \frac{x_{o}^{2}}{B_{o}\eta_{H}} = \frac{x_{o}}{\frac{B_{o}\eta_{H}}{x_{o}}}.
 \end{equation}
It is clear that the denominator of the last fraction corresponds to some velocity. With this is defined a new velocity
 \begin{equation}
    \lambda_{H} = u + c_{f} + C\frac{B_{o}\eta_{H}}{x_{o}}, 
 \end{equation}
 \label{eq.lambdaHall}
where $u$ is the fluid velocity and $c_{f}$ is the fast magnetosonic velocity. According to \cite{Toth2008}, it is not necessary to include the exact speed of the modes introduced by the Hall effect, but a reduced speed can be used to guarantee the stability of the simulations. In this case, $C$ is a constant that allows defining a reduced or extended speed of the Hall term. Thus, there is freedom in the choice of $C$ that allows to better adapt the time step depending on the physical system to be simulated.

At each time level MAGNUS evaluates the speed $\lambda_{H}$ at all points in the domain and chooses the largest. Then, the time step is calculated as
\begin{equation}
    \Delta t_{H} = C_{CFL} \times \left( \frac{\Delta x}{\lambda_{H\ i,j,k}^{n, x}} + \frac{\Delta y}{\lambda_{H\ i,j,k}^{n, y}} + \frac{\Delta z}{\lambda_{H\ i,j,k}^{n, z}} \right).
    \label{dt.hall}
\end{equation}
The same is done with the ambipolar term, for which
\begin{equation}
    \lambda_{A} = u + c_{f} + \frac{B_{o}^{2}\eta_{A}}{x_{o}}, 
\end{equation}
 \label{eq.lambdaAmb}
and
\begin{equation}
    \Delta t_{A} = C_{CFL} \times \left( \frac{\Delta x}{\lambda_{A\ i,j,k}^{n, x}} + \frac{\Delta y}{\lambda_{A\ i,j,k}^{n, y}} + \frac{\Delta z}{\lambda_{A\ i,j,k}^{n, z}} \right).
    \label{dt.ambipolar}
\end{equation}
In the end, MAGNUS chooses the smallest time step between (\ref{dt.resistive}), (\ref{dt.hall}) and (\ref{dt.ambipolar}).

\subsection{Scaling and Nondimensionalization}

As mentioned above, MAGNUS solves the dimensionless system of equations. Therefore, the dimensionless equations are the ones that evolve. Then the dimensionless quantities that result from the system's evolution can be scaled with quantities that do have dimensions and depend on the simulated physical system.

In the code, this is done by dividing each physical quantity by characteristic quantities of the system such as $l_{a}$ $\rho_{a}$, $u_{a}$ and constants such as the magnetic permeability $\mu_{0}$. Including the Hall and ambipolar terms in the code makes it necessary to determine expressions for $\eta_{Ha}$ and $\eta_{Aa}$.

\section{GEM reconnection challenge}
\label{sec:test}

After the implementation of the Hall and ambipolar term in the MAGNUS code was done, we proceeded with the Geospace Environmental Modeling (GEM) reconnection challenge \citep{GEM2001}. The analytical solution to this problem is not known, but since its publication, several authors have reproduced their results, which is why it is often used as a benchmark test \citep{Toth2008, Strumik2017}.

First, we use the GEM as a numerical test to verify the implementation of the Hall term in MAGNUS. For the test, the system starts from rest with a magnetic field whose components are defined as
\begin{gather}
B_{x} = \tanh(2z) + \delta B_{x}, \\
B_{z} = \delta B_{z}, \\
B_{y} = 0.
\end{gather}
The perturbations $\delta B_{x}$ and $\delta B_{z}$ are defined as
\begin{gather}
\delta B_{x} = \frac{0.05\pi}{L_{z}} \cos\left(\frac{2\pi x}{2L_{x}}\right) \sin\left(\frac{\pi z}{2L_{z}}\right), \\
\delta B_{z} = \frac{0.1 \pi}{L_{x}} \sin\left(\frac{2\pi x}{2L_{x}}\right) \cos\left(\frac{\pi z}{2L_{z}}\right).
\end{gather}

The initial density is 
\begin{equation}
\rho = 1.2 - [\tanh(2z)]^{2},
\end{equation}
and the pressure is $p=\rho/2$. In all simulations we set the resistivity to $\eta=0.005$. In the resistive+Hall simulations, the  normalized ion mass per charge is set to  $1.0$, so that $\eta_ {H}=1/\rho$.

In both cases, resistive and resistive+Hall, the simulation box is $[-L_{x},L_{x}] \times [-L_{z},L_{z}] \times [-L_{y},L_{y}]$, where $L_ {x}=12.8$, $L_{z}=6.4$ and $L_{y}=0.2$. In this volume, the equations are discretized in a three-dimensional numerical grid with $N_{x}=64$, $N_{z}=128$ and $N_{y}=4$. The boundary conditions are periodic in $x$, inflow in $z$ and outflow in $y$.

\begin{figure}[h]
\begin{center}
\noindent\includegraphics[width=0.48\textwidth]{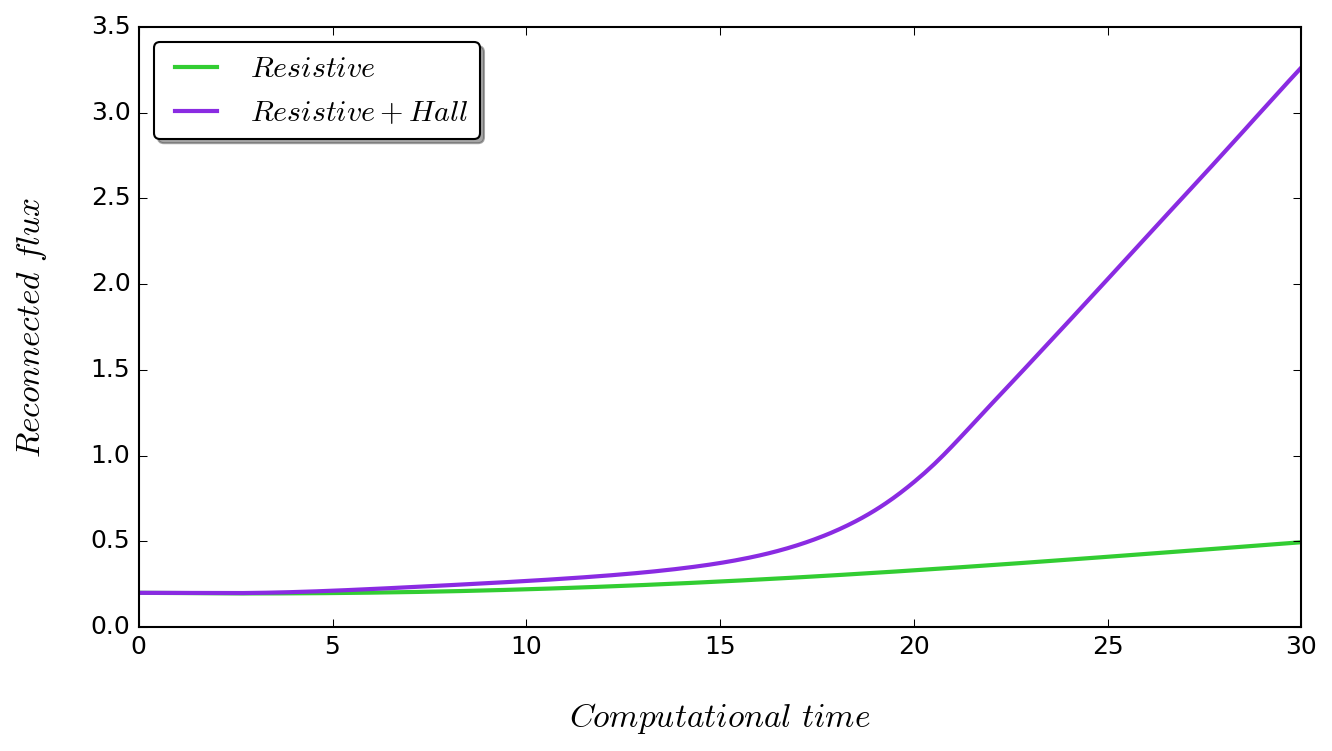}
\caption{Testing results of the reconnected flux for the initial data of the GEM reconnection challenge.}
\label{fig1}
\end{center}
\end{figure}

Figure \ref{fig1} shows the reconnected flux of resistive and resistive+Hall simulations. In both, the flux is calculated at each time step as
\begin{equation}
\textnormal{Reconnected flux} = \int_{N_{x}/2}^{N_{x}} B_{z}(x,N_{y}/2,N_{z}/2)dx.
\end{equation}
With the Hall effect, the reconnected flux reaches a value of $3.3$, about six times larger than in the simulation with resistivity alone, where the reconnected flux reaches a value of $0.5$. This result is in qualitative and quantitative agreement with those presented in GEM for MHD and Particle-In-Cell (PIC) simulations. For comparison see Figure 1 of \cite{GEM2001}, at $t=30$ the Hall MHD case reaches a reconnected flux of about $3.2$. See the bottom panel of Figure 1 from \cite{GEM_Pritchett2001}, the reconnected flux reaches a value slightly above $3.0$.

These results are also consistent with those of other authors who have used GEM as a benchmark. For example, in the kinetic simulation of \cite{Schmitz2006} the curve reaches a value slightly above $3.0$, similar to the MHD simulation of \cite{Toth2008}, both at time $t=30$. This differs slightly from the results of \cite{Strumik2017}, whose MHD simulation reaches a value of $2.5$, lower than the simulations mentioned above and the results shown in Figure \ref{fig1}. For the resistive case, the result presented in Figure \ref{fig1} is also in agreement with those presented by \cite{GEM2001, Toth2008, Strumik2017}.

\section{Ambipolar diffusion in the GEM}
\label{sec:ambipolarGEM}

Although the GEM was designed to study the Hall effect on magnetic reconnection, we use the same scenario to study the effect of ambipolar diffusion. To our knowledge, such a systematic study of simulations with ambipolar diffusion in the GEM model has not been done before. In fact, ambipolar diffusion had been overlooked in the Earth's magnetotail. %Recently, however, evidence from the Magnetospheric Multi-Scale (MMS) mission has come to highlight the importance of transverse ambipolar electric fields in the structure and dynamics of current sheets \citep{dubois2022}.

We present two types of simulations: resistive+ambipolar and resistive+Hall+ambipolar. The Hall parameter remains unchanged, with the same value as in the previous section ($\eta_ {H}=1/\rho$). The ambipolar term, $\eta_{A} = K_{A} \cdot 1/\rho^{2}\sqrt{T}$, is implemented with $T=p/\rho$. Three values of $K_{A}$ are considered: $0.001$, $0.005$, and $0.01$ (since $K_{A}$ is inversely proportional to the degree of ionization, the latter value represents the less ionized case). These values were selected taking into account equation (22) from \citep{Ni2015}.
These values physically correspond to values of $\eta_{A}$ in the chromosphere. The code being dimensional allows for the straightforward use of these values without any issues.

\begin{figure}[h]
\begin{center}
\noindent\includegraphics[width=0.48\textwidth]{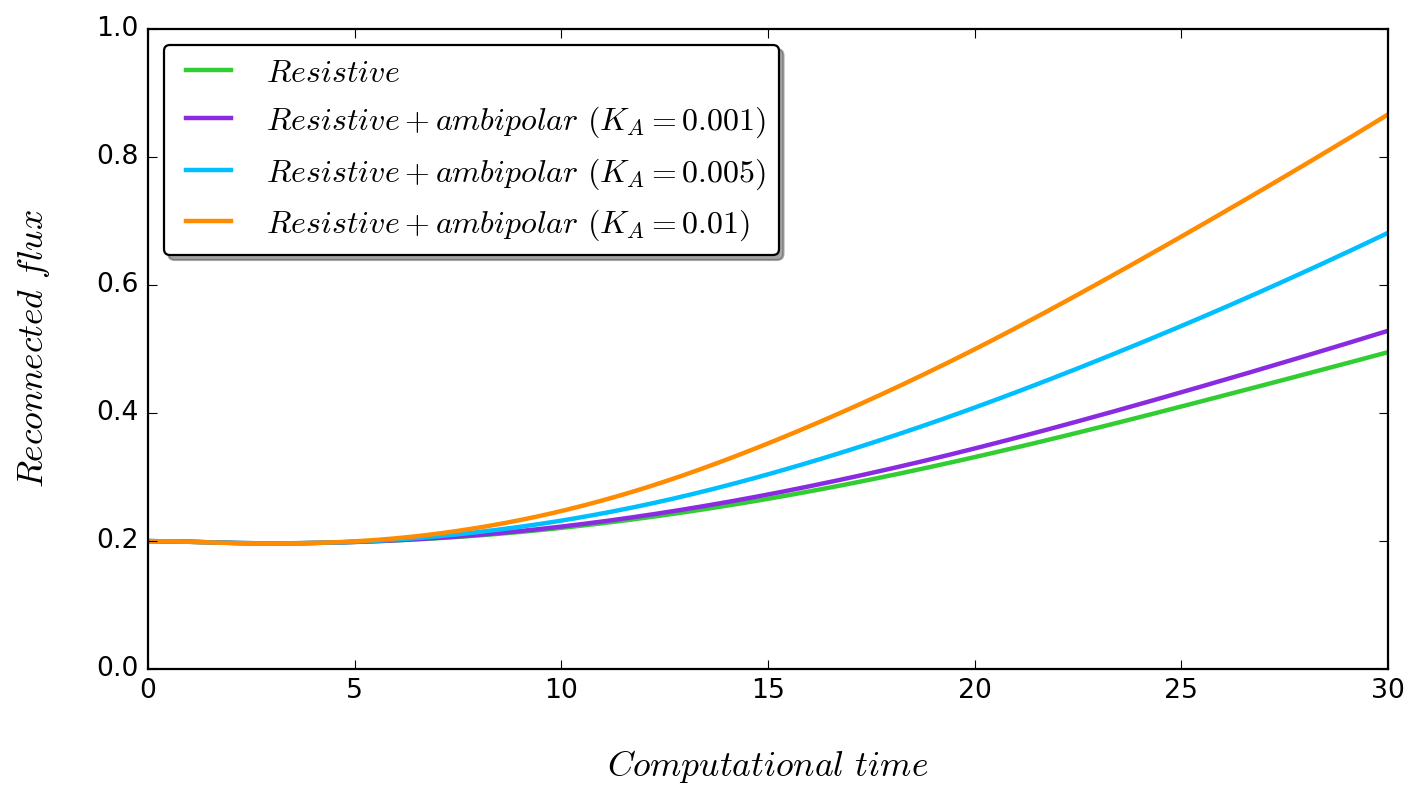}
\caption{Reconnected flux for different values of the $K_{A}$ parameter in the resistive+ambipolar simulations.}
\label{fig2}
\end{center}
\end{figure}

\begin{figure}[h]
\begin{center}
\noindent\includegraphics[width=0.48\textwidth]{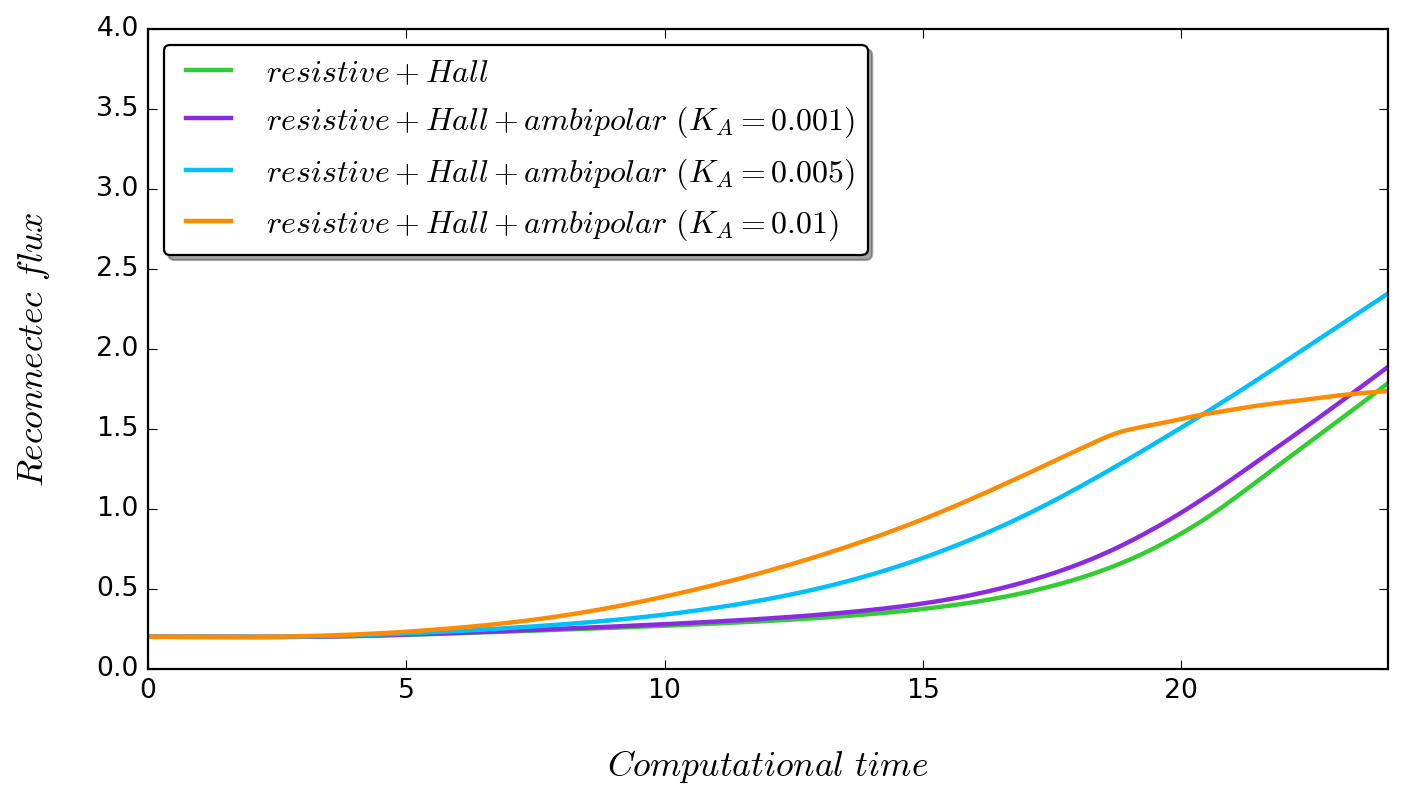}
\caption{Reconnected flux for different values of the $K_{A}$ parameter in the resistive+Hall+ambipolar MHD simulations.}
\label{fig3}
\end{center}
\end{figure}

Figure \ref{fig2} shows the reconnected flux of the resistive simulation compared to the resistive+ambipolar cases. As can be seen, the reconnected flux grows with the value of $K_{A}$, obtaining an increase of about $7\%$ for $K_{A}=0.001$, $38\%$ for $K_{A}=0.005$, and $75\%$ for $K_{A}=0.01$, all percentages with respect to the resistive case. Comparing with figure \ref{fig1}, we see that the increase of the reconnected flux is much higher with the Hall effect than with the ambipolar diffusion, but the percentages indicate that the ambipolar diffusion also contributes to the reconnected flux. Furthermore, its increase could be higher if other values of $K_{A}$ are considered. Figure \ref{fig3} compares the reconnected flux of the resistive+Hall simulation with the resistive+Hall+ambipolar cases. 
%\textbf{For $K_{A}=0.001$ and $K_{A}=0.005$, the combination of the Hall effect and ambipolar diffusion causes an anomalous growth of the reconnected flux (see purple and blue curves at times $t=27$ and $t=24$, respectively). Nevertheless, in both cases, the presence of the ambipolar effect increases the reconnected flux by $4\%$ and $32\%$ at that time.}
For the case with $K_{A} = 0.01$, the reconnected flux grows faster until it reaches an increase of $143\%$ (with respect to the resistive+Hall case) at time $t=18$ and then remains approximately stationary until the end. This quasi-steady state of the reconnected flux seems to be caused by the appearance of a plasmoid in the centre of the domain at $t=18$.

\begin{figure}[h]
\begin{center}
\noindent\includegraphics[width=0.48\textwidth]{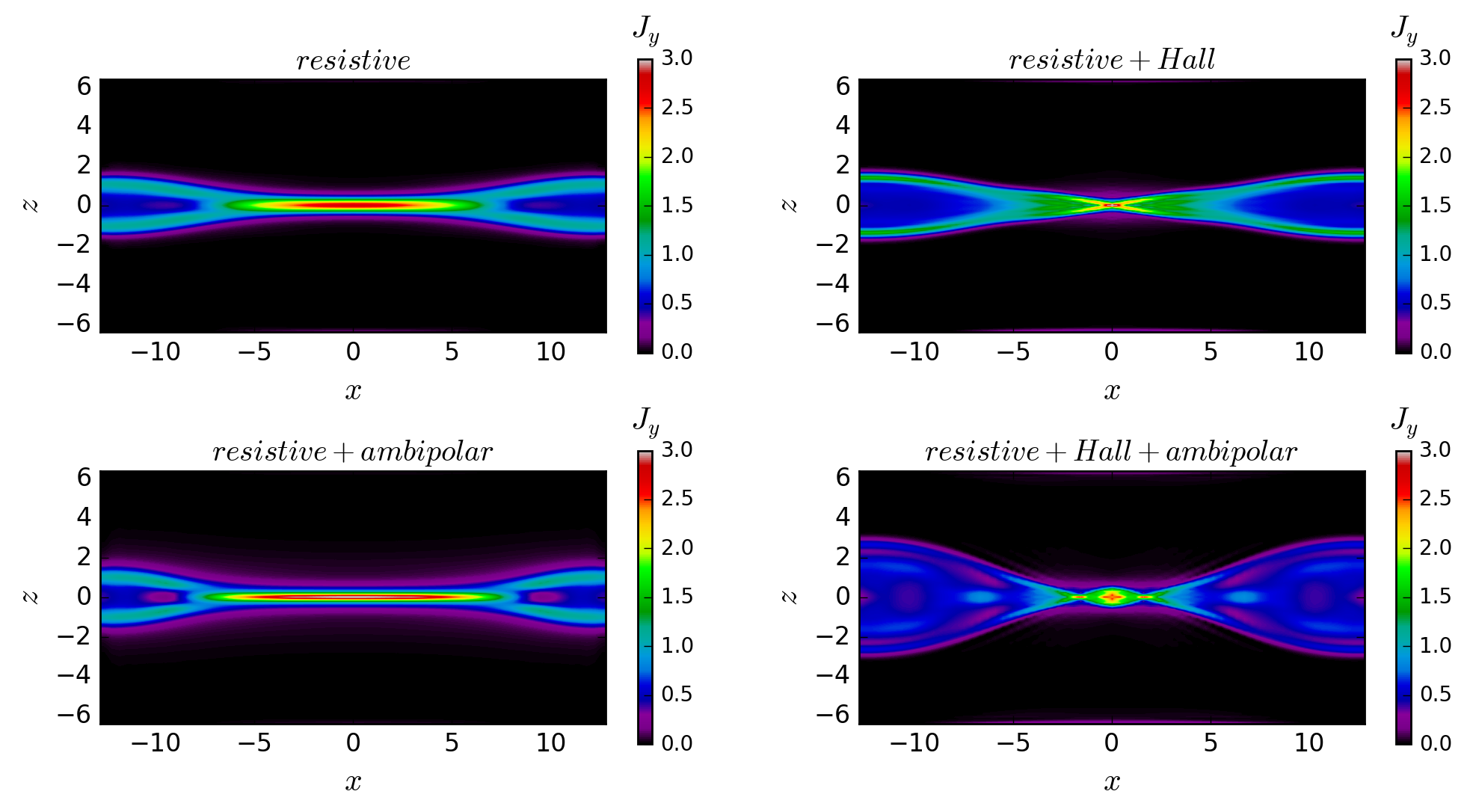}
\caption{Out-of-plane component of the current density at time $t=20$. The simulations with Hall effect and ambipolar diffusion have parameters $K_{H}=1.0$ and $K_{A}=0.01$.}
\label{fig4}
\end{center}
\end{figure}

%\textbf{For $K_{A}=0.001$ and $K_{A}=0.005$, the combination of Hall and ambipolar effects causes the current sheet to inflate at time $t=28$ and $t=25$, respectively. Then, there is a kind of shearing that causes the current sheet to break. The inflation occurs at the same time that the reconnected flux curves of Figure \ref{fig3} start to have an anomalous growth. Thus, we found that at least for these two values of our $K_{A}$ parameter, there are time scales in which the flow increases notably, but after a certain time the system inflates, and although the reconnected flux continues to increase, this result is not physical.} %since is the opposite of the typical scenario: where the reconnection rate increases since the ambipolar diffusion makes the current sheet thinner.
To illustrate the qualitative differences between some of the simulations carried out, Figure \ref{fig4} shows the current density at time $t=20$ for the four types of simulations: resistive, resistive+Hall, resistive+ambipolar, and resistive+Hall+ambipolar, using $K_{A}=0.01$ for those that include the ambipolar term. For the resistive simulation (upper left panel), we observe an elongated current sheet of nearly the order of the global length scale, as proposed by the model of Sweet-Parker. By including the ambipolar term (lower left panel) we observe a thinner current sheet, in agreement with the results of \cite{Ni2015}, where the inclusion of ambipolar diffusion causes a rapid thinning of the current sheet in the solar chromosphere.
%Since our simulations are performed in the GEM scenario, ambipolar diffusion may also be a mechanism for current sheet thinning in the magnetosphere, and not only in plasmas such as the solar chromosphere or molecular clouds.

Together with the reconnected flux shown in Figure \ref{fig2}, these results support the fact that elongated current sheets result in low reconnection rates, which are inconsistent with the rate suggested by observations. For the resistive+Hall simulation (upper right panel), the diffusion region is much smaller as the current density is concentrated in an x-shaped region in the centre of the domain, a result consistent with the one presented by \cite{Strumik2017}. However, it differs from the presented by \cite{Toth2008}, whose simulation with the Hall effect shows asymmetric behaviour, has two reconnection regions and a plasmoid, a result that is not obtained with MAGNUS for any time of the resistive+Hall simulation case. In fact, the combination of the Hall effect with ambipolar diffusion in the resistive+Hall+ambipolar case is the only one that produces the formation of a plasmoid (bottom right panel). %WHY?????

\begin{figure}[h]
\begin{center}
\noindent\includegraphics[width=0.48\textwidth]{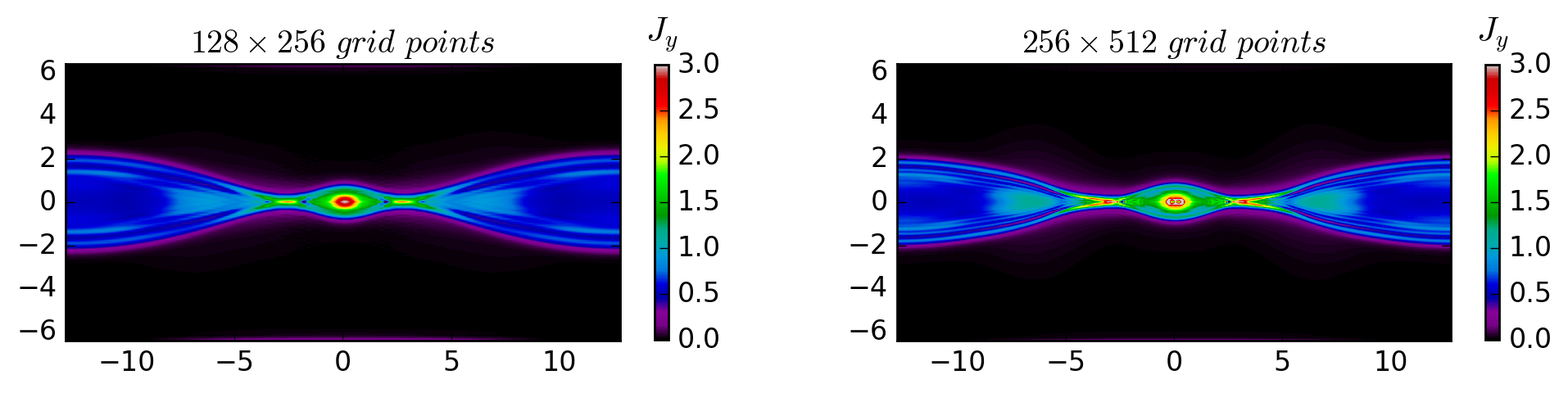}
\caption{Out-of-plane component of the current density at time $t=18$ with higher resolutions.}
\label{plasmoid_res}
\end{center}
\end{figure}

It is worth checking whether the plasmoid is due to insufficient resolution. For that, in Figure \ref{plasmoid_res} we present the results of two more simulations with higher resolution: the first one with $128 \times 256$ and the second with $256 \times 512$ grid points. Both panels of Figure \ref{plasmoid_res} show the same global morphology. Locally, the higher resolution panel reveals more detail, but the appearance of the plasmoid in the centre of the domain is consistent in simulations with different resolutions. Based on these results, we can confirm that the appearance of the plasmoid is not due to a numerical issue.

Concerning the formation of the plasmoid, we want to highlight the current sheet thinning caused by the ambipolar diffusion that can be seen in \ref{fig4}. We believe that the spatial scale reduction, attributed to the thinning of the current sheet characteristic of ambipolar diffusion, is likely crucial for bringing magnetic field lines close enough together to facilitate magnetic reconnection. Moreover, in a region of strong magnetic gradient, ambipolar diffusion can allow electrons to move faster than ions, which can cause the formation of thin current layers and then the possible appearance of instabilities in those layers. These instabilities can evolve towards the formation of plasmoids.

We believe that the formation of the plasmoid in the simulation is facilitated by the combined effects. However, substantiating this claim requires a systematic study solely focused on plasmoid formation, demonstrating how these combined effects enhance the likelihood of plasmoid formation. It is crucial to emphasize that plasmoid formation cannot be attributed to a single phenomenon. Rather, it emerges from the interaction and dynamics of multiple factors. As noted by \cite{singh2019effect}, while the thinning of the current sheet plays an important role in plasmoid formation, it is influenced not only by this thinning but also by additional factors that collectively shape the overall dynamics of magnetic reconnection.

\section{Current sheet with a guide-field}
\label{sec:cs}

This second model attempts to simulate the process of magnetic reconnection from the formation of a current sheet. It also includes the presence of a guide field, that is, the non-reconnecting component of the magnetic field that is out-of-plane. This out-of-plane component is commonly used in solar flare simulations and plays an important role in determining key properties of reconnection, like the efficiency of particle acceleration \citep{dahlin2022}.
Unlike the GEM, in this model, we use initial data with constant density, constant pressure, and magnetic field without perturbations. This ensures that the current sheet is not formed at time zero and that the reconnected flux is zero. To trigger reconnection, we use spatially localized resistivity.

The system starts from rest with a magnetic field whose components are defined as \citep{Shibata2022}
\begin{gather}
B_{x} = B \tanh(y/w), \\
B_{y} = 0, \\
B_{z} = B / \cosh(y/w), 
\end{gather}
being $B_{z}$ the guide-field with $B=3.92$ and $w=0.5$. Density and initial pressure of the system are given by $\rho=1.0$ and $p=1.0$, and the localized resistivity is given by the function
\begin{equation}
\eta = \eta_0 \cdot \exp \left[-\left(\frac{\sqrt{y^2+\left(x-h_\eta\right)^2}}{w_\eta}\right)^2\right],
\label{eq.resistivity}
\end{equation}
with $\eta_{0}=0.01$, $h_\eta=15.0$ and $w_\eta=1.0$. 
%For the ambipolar coefficient, we used $K_{A}=0.01$ and for the Hall coefficient, we set the normalized ion mass per charge to $0.2$.
The computational domain is $[0,L_{x}] \times [-L_{y}/2,L_{y}/2] \times [-L_{z}/2,L_{z}/2]$ where $L_{x}=30.0$, $L_{y}=15.0$ and $L_{z}=0.15$, with a numerical mesh with $N_{x}=800$, $N_{y}=400$ and $N_{z}=4$. The boundary conditions are periodic in $x$, inflow in $y$ and outflow in $z$.

With the previous specifications, four simulations are performed: resistive, resistive+Hall, resistive+ambipolar, and resistive+Hall+ambipolar. It is important to emphasize that in order to perform systematic simulations, the only thing that varies in each type of simulation is the presence or absence of the Hall effect and the ambipolar diffusion, which is controlled by the parameters $\eta_{H}$ and $\eta_{A}$. The initial data, boundary conditions, computational domain, spatial resolution, and numerical methods are the same for all simulations. For the ambipolar coefficient, we used $K_{A}=0.01$ and for the Hall coefficient, we set the normalized ion mass per charge to $0.2$. 

As mentioned above, the out-of-plane component of this model is also commonly used in solar flare simulations. That is one of the reasons why this second model is suitable for application to partially ionized plasmas, like the chromosphere, where ambipolar diffusion is a relevant phenomenon and could potentially impact its dynamics. In such instances, it is essential to select characteristic values for scaling the dimensionless outcomes derived from the computational code, so that the results may have dimensions according to the physical scenario. In our study, all quantities were kept dimensionless, meaning they were not scaled to fit a particular scenario. This approach allowed us to focus on assessing the potential impact and effects of the Hall and ambipolar diffusion terms on reconnection itself, rather than their applicability to a specific context.

\subsection{Current sheet morphology}

\begin{figure}[h]
\begin{center}
\noindent\includegraphics[width=0.48\textwidth]{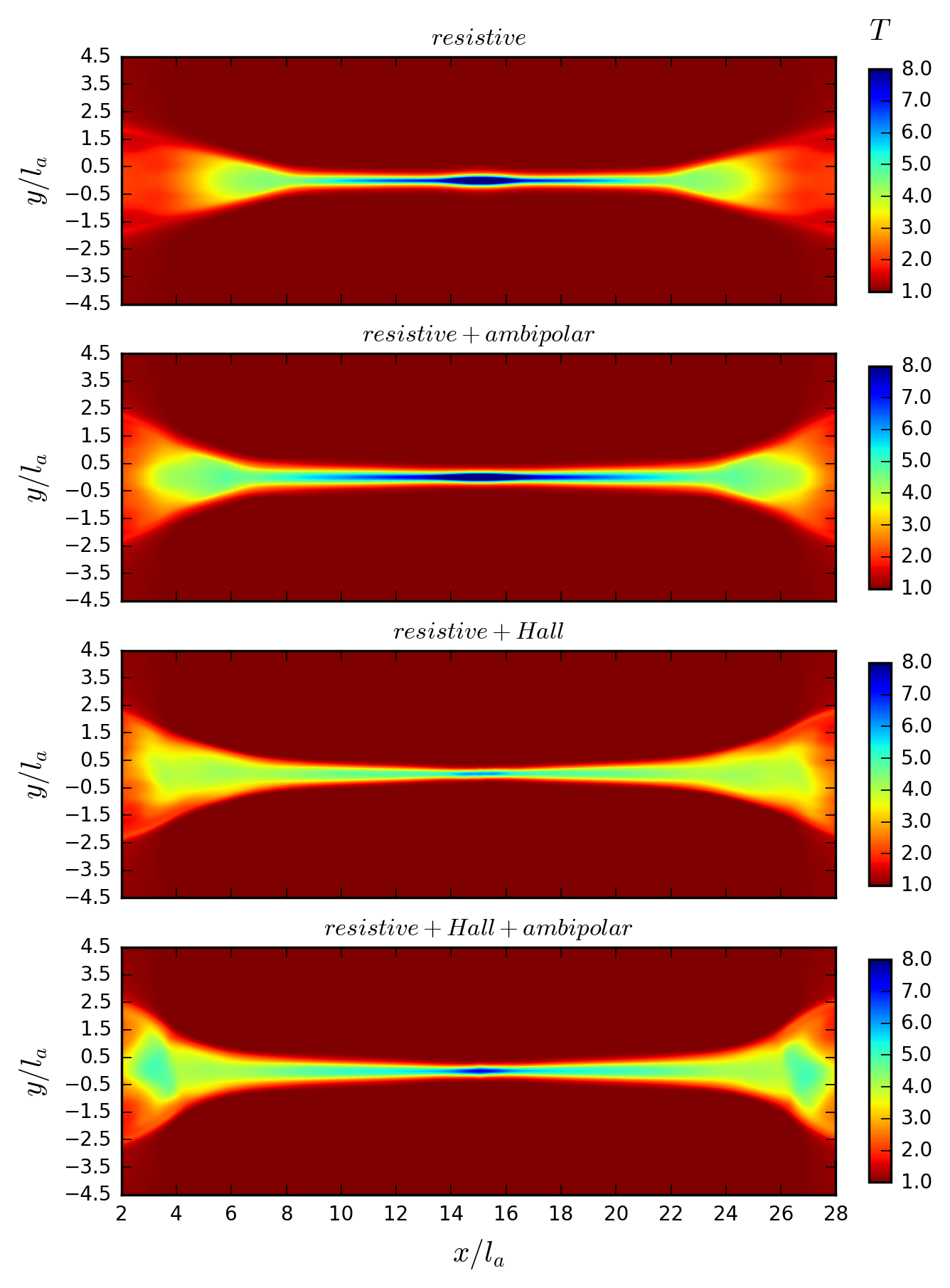}
\caption{Temperature distribution at $t=12t_{a}$. In all cases, the hottest region is found at the current sheet centre. For the resistive+Hall+ambipolar simulation, high heating is also observed at the extremes of the domain.}
\label{fig7}
\end{center}
\end{figure}

First, we examine the current sheet morphology at time $t=12t_{a}$, where $t_{a}$ is a characteristic time. Figure \ref{fig7} shows the temperature distribution in the current sheet for each simulation. It can be seen that the current sheet centre is the hottest region, especially in the resistive and resistive+ambipolar cases. For the resistive+Hall and resistive+Hall+ambipolar simulations, the central region is not as hot, but we see heating at the extremes of the domain, particularly in the presence of the ambipolar term.

\begin{figure}[h]
\begin{center}
\noindent\includegraphics[width=0.48\textwidth]{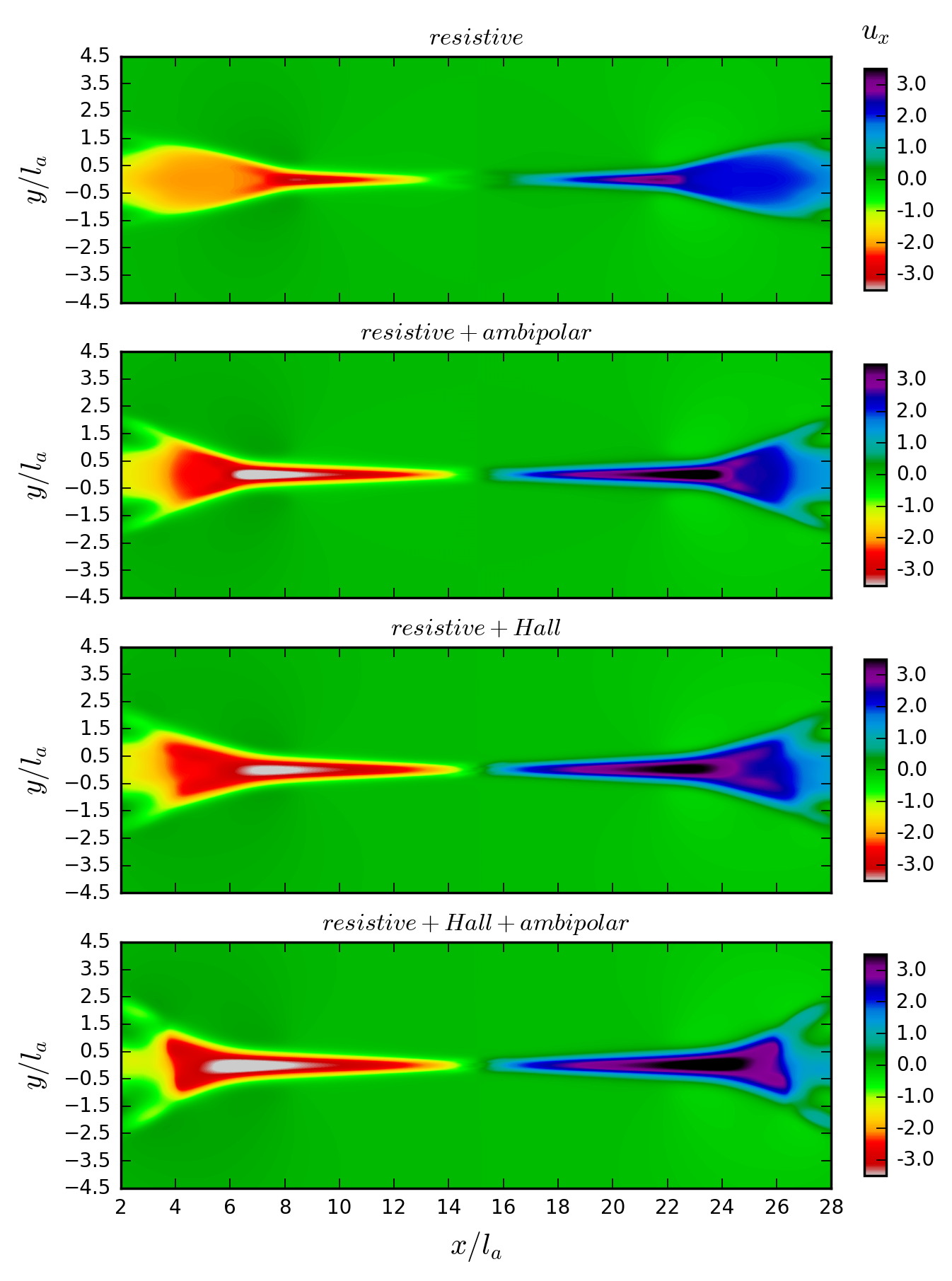}
\caption{$u_{x}$ velocity maps at $t=12t_{a}$. All simulations exhibit two flows with opposite directions, a characteristic feature of magnetic reconnection. Moreover, the highest velocity regions widen in the presence of the Hall effect and ambipolar diffusion.}
\label{fig8}
\end{center}
\end{figure}

Regarding the plasma velocities, Figure \ref{fig8} shows maps for the $u_{x}$ component, where we can clearly see the presence of two flows with opposite directions coming out of the central region. This indicates that the plasma is being accelerated out of the current sheet, one of the main consequences of reconnection. Therefore, this type of velocity diagram is characteristic of magnetic reconnection and is also consistent with the diagrams in the PIC simulations of \cite{Nakamura2018}, which were performed to model a reconnection event detected by MMS. The different panels of figure \ref{fig8} also show that the high-velocity regions are bigger in the presence of the Hall effect and ambipolar diffusion. Thus, both effects contribute to the acceleration of particles.

\begin{figure}[h]
\begin{center}
\noindent\includegraphics[width=0.48\textwidth]{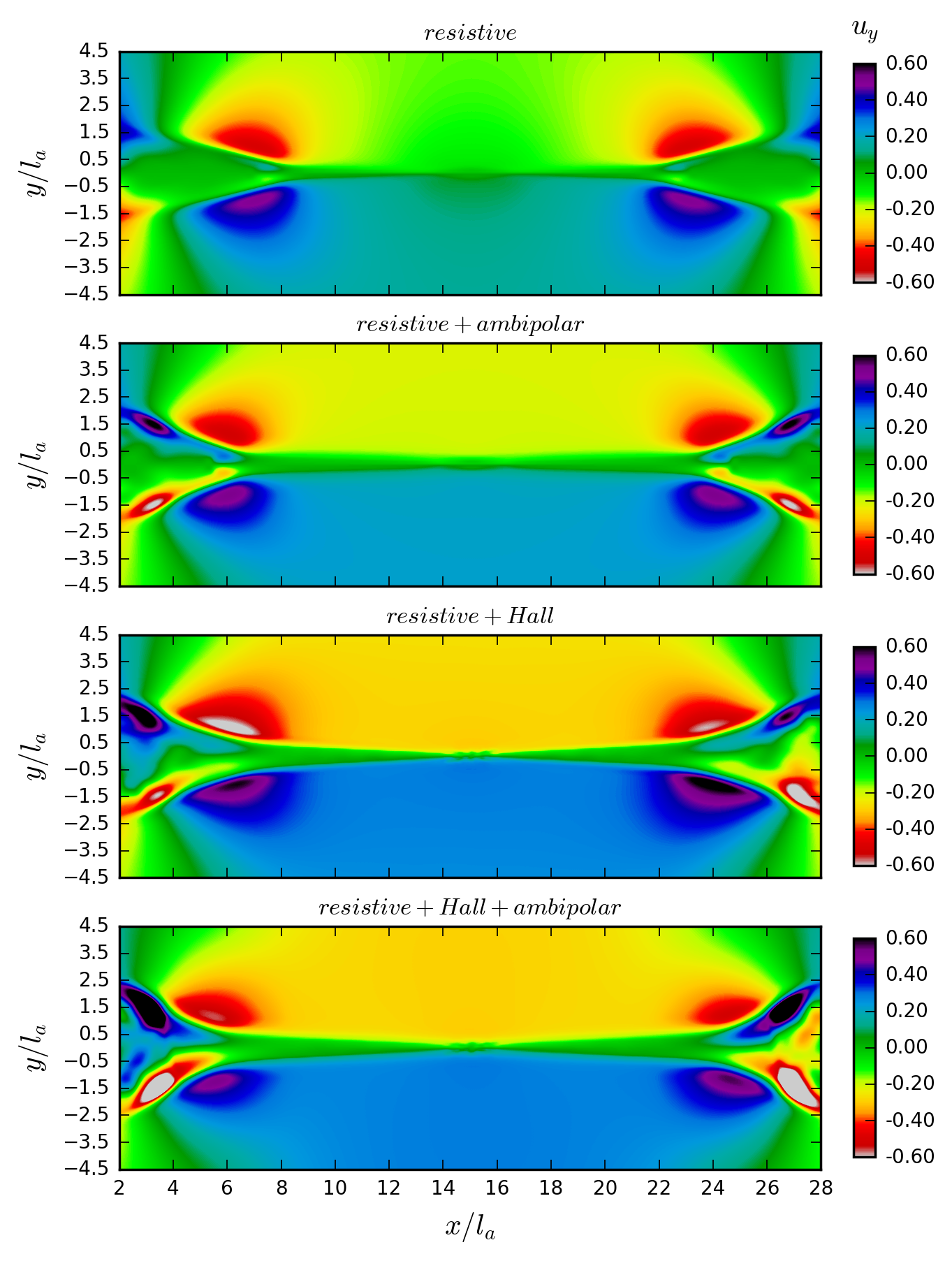}
\caption{$u_{y}$ velocity maps at $t=12t_{a}$. All maps show plasma movement towards the current sheet and velocity distribution changes in its extremes. This indicates plasma is entering and exiting the diffusion region. Regions with higher velocities are found at the current sheet extremes and are highest in the presence of the Hall effect and ambipolar diffusion.}
\label{fig9}
\end{center}
\end{figure}

Finally, figure \ref{fig9} shows maps for the $u_{y}$ component. In the panels, we see that the plasma is moving toward the current sheet, with positive velocities in the lower region of the domain and negative velocities in the upper region. Towards the extremes of the domain, where the current sheet opens, this distribution of velocities changes: negative velocities are present in the lower region and positive in the upper region. This indicates that although the plasma enters the diffusion region, it is also expelled from it according to the scheme of magnetic reconnection. We can also see that the plasma regions with the highest velocities are located at the extremes of the current sheet. There, the colour distribution indicates that there are abrupt changes in both the magnitude and direction of the velocity. This type of behaviour can trigger turbulence and be another mechanism contributing to plasma acceleration \citep{Price2016}. Three-dimensional MHD simulations of \cite{Shibata2022} and \cite{shen2022origin} investigate the dynamics of solar flares with a focus on the generation of turbulence. These studies identified regions characterized by turbulent flows, which may significantly influence electron acceleration within these areas. Additionally, they found self-organized structures in turbulent interface flare regions, resembling formations also seen in supernova remnants. Further investigation on this topic is essential for understanding the mechanisms behind substantial energy releases and events of particle acceleration.

\subsection{Reconnection rate}

%As mentioned above, the reconnection rate can control the amplitude of geomagnetic disturbances.
In addition to telling us that the flux is being reconnected, the reconnection rate is associated with the energy released during the process.

The original model for determining the reconnection rate was proposed by \cite{Sweet1958a, Sweet1958b} and \cite{Parker1957, Parker1963}, considering a diffusion region with length L, on the order of the global external length scale that occupies the entire boundary between opposing magnetic fields. According to this model, the dimensionless reconnection rate is given by: $v_{i}/v_{Ai}=1/\sqrt{S}$, where $v_{i}$ is the characteristic speed of entry of the field lines into the diffusion region, $v_{Ai}$ is the Alfvén speed at the entry and $S$ is the Lundquist number. In general, the Lundquist number is large for most astrophysical plasmas ($S>>10^{6}$). Consequently, the Sweet-Parker reconnection is too slow to explain phenomena such as geomagnetic storms \citep{Birn2007}.

Simulations and observations have reported a fast rate of 0.1, and smaller values cannot explain the rapid energy release that accelerates the plasma. Here the dimensionless reconnection rate is calculated as $v/v_{A}$, where $v$ is the plasma velocity at the entrance of the diffusion region and $v_{A}$ is the Alfvén velocity, also at the entrance \citep{Priest2000}. Although it is impossible to determine the exact volume and location of the diffusion region, the localized resistivity in the centre of the region implies that we have a single reconnection site also located in the centre. Based on this, the entry to the diffusion region can be considered as the point located at ($N_{x}/2,N_{yi},N_{z}/2$), where $N_{yi}$ is a point close to the line $y=0.0$ ($N_{y}/2$), but not on it because there the magnetic field is zero. Thus, for all simulations, the reconnection rate is calculated as
\begin{equation}
    \textnormal{Reconnection rate} = \frac{v(N_{x}/2,N_{yi},N_{z}/2)}{v_{A}(N_{x}/2,N_{yi},N_{z}/2)},
\end{equation}
with $N_{yi}$ on the line $y=0.5$. The velocities are computed using only the components of the $xy$ plane, that is, $v=\sqrt{v_{x}^{2}+v_{y}^{2}}$ and $v_{A}=B/\sqrt{\rho}$ with $B=\sqrt{B_{x}^{2}+B_{y}^{2}}$.

\begin{figure}[h]
\begin{center}
\noindent\includegraphics[width=0.48\textwidth]{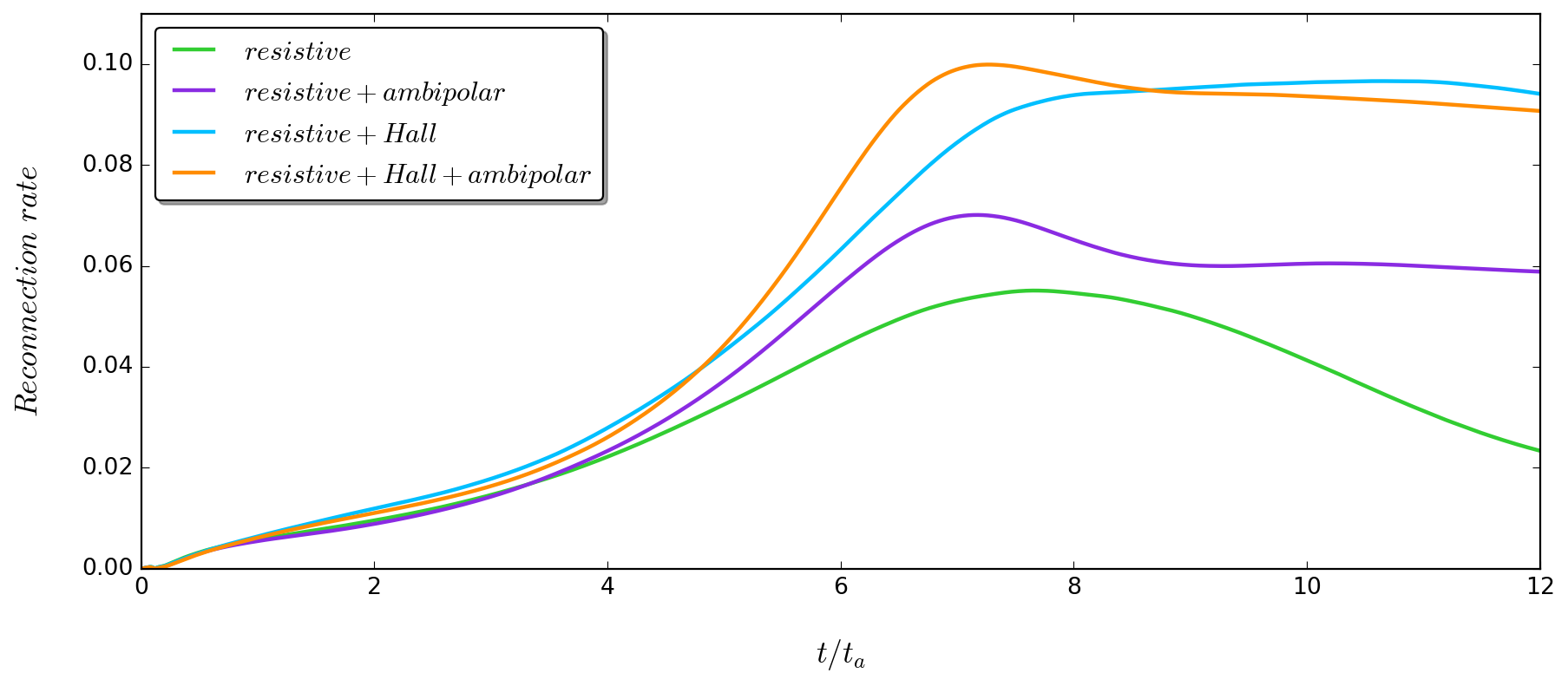}
\caption{Reconnection rate as a function of time.}
\label{fig_10}
\end{center}
\end{figure}

In Figure \ref{fig_10} we plot the reconnection rate as a function of time for the four types of simulations. The results show that when running systematic simulations, the resistive+Hall+ambipolar case reaches a reconnection rate of $0.1$, followed by the resistive+Hall case. The Hall effect simulations show similar reconnection rates, implying that the Hall term is indeed important to obtain reconnection rates of $0.1$. However, the results also show that the ambipolar term may play an important role in the reconnection rate.  While the Hall effect dominates when reconnection rates are compared to the resistive case, it is the combination of the Hall effect and ambipolar diffusion that reaches a value of $0.1$, comparable to what has been observed, for example, in the magnetosphere, where MMS values are $0.1-0.2$ \citep{Chen2017, Nakamura2018}.

The ambipolar term also seems to influence the reconnection rate to increase faster, which is why the maximum rate appears earlier in the simulations with ambipolar diffusion, an effect also reported by other authors (see case C from model I and case F from model II in Figure 5 of \cite{Ni2015}). However, \cite{Ni2015} attribute this to the fact that ambipolar diffusion triggers plasmoid instability. Our results show that even without plasmoids, ambipolar diffusion has the effect of accelerating the reconnection process.

In our simulations, the Lundquist number is of the order of $S \sim 10^4$, which corresponds to a reconnection rate $S^{-1/2} \sim 0.01$ according to Sweet-Parker (as we mentioned above, the rate is calculated as $1/\sqrt{S}$). This shows that reconnection in our simulations is not a Sweet-Parker-type reconnection, which is not surprising since the Sweet-Parker model involves several simplifications. For example, it assumes a uniform resistivity, contrary to the localized resistivity we used for our simulations (as described in equation \eqref{eq.resistivity}).

Finally, concerning the formation of plasmoids, it is worth mentioning that for values of $S > 10^4$, plasmoid instability occurs \cite{loureiro2007instability}. In the solar chromosphere, for instance, the values of the Lundquist number are of the order of $S \sim 10^6 - 10^8$ \cite{Ni2015}. Because the value we assume in our work was the critical value $S \sim 10^4$, we do not expect the formation of plasmoids.

\subsection{Energy conversion and transport}

\begin{figure}[h]
\begin{center}
\noindent\includegraphics[width=0.48\textwidth]{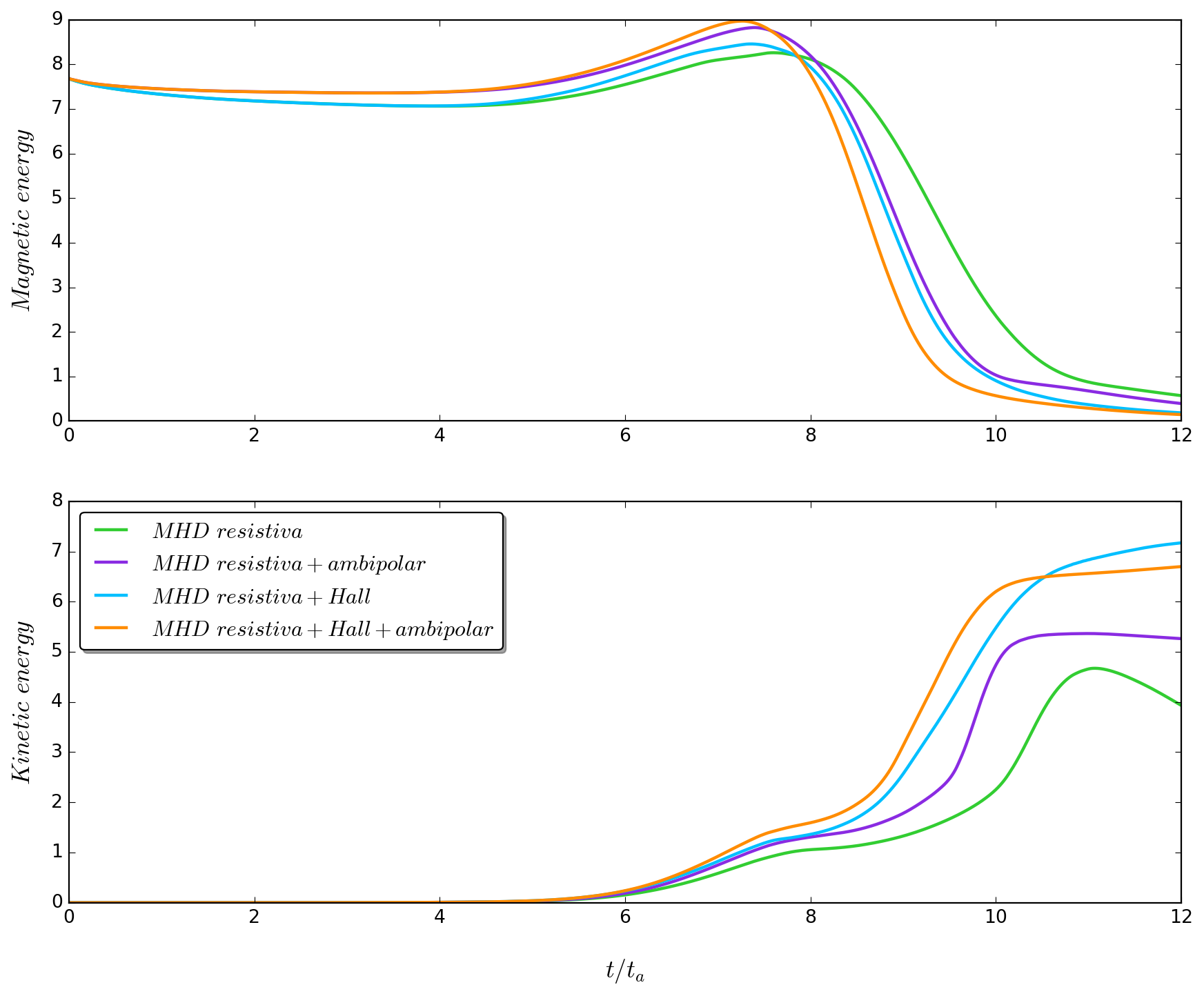}
\caption{Temporal evolution of kinetic and magnetic energies.}
\label{fig_11}
\end{center}
\end{figure}

Figure \ref{fig_11} shows the evolution of magnetic and kinetic energy measured by a detector. Similar to the reconnection rate, this detector corresponds to a point with coordinates ($N_{xi},N_{y}/2,N_{z}/2$), where $N_{xi}$ is at $x=10.0$. The results show that the kinetic energy starts from zero for the four types of simulations, in agreement with the initial data. From the initial time to $t=5t_{a}$, there is a slight decrease in the magnetic energy that does not lead to a noticeable increase in the kinetic energy, most likely because there is also conversion to internal energy and dissipation by Joule heating. From time $t=5t_{a}$ the kinetic energy increases, coinciding with an increase in magnetic energy from $t=5t_{a}$ to $t=7t_{a}$. Finally, from $t=7t_{a}$ to the end of the simulation, the behaviour of the curves is consistent with the reconnection process: the magnetic energy decreases while the kinetic energy increases because there is a conversion from one to the other.

The resistive+Hall and resistive+Hall+ambipolar simulations achieve the highest kinetic energy values. This suggests that the Hall and ambipolar terms not only affect the reconnection rate but also the particle acceleration. Considering that the kinetic energy of the resistive+Hall and resistive+Hall+ambipolar simulations reaches a value close to the initial value of the magnetic energy, we conclude that the Hall effect and its combination with ambipolar diffusion provide a more efficient energy conversion during the reconnection process.

\begin{figure}[h]
\begin{center}
\noindent\includegraphics[width=0.48\textwidth]{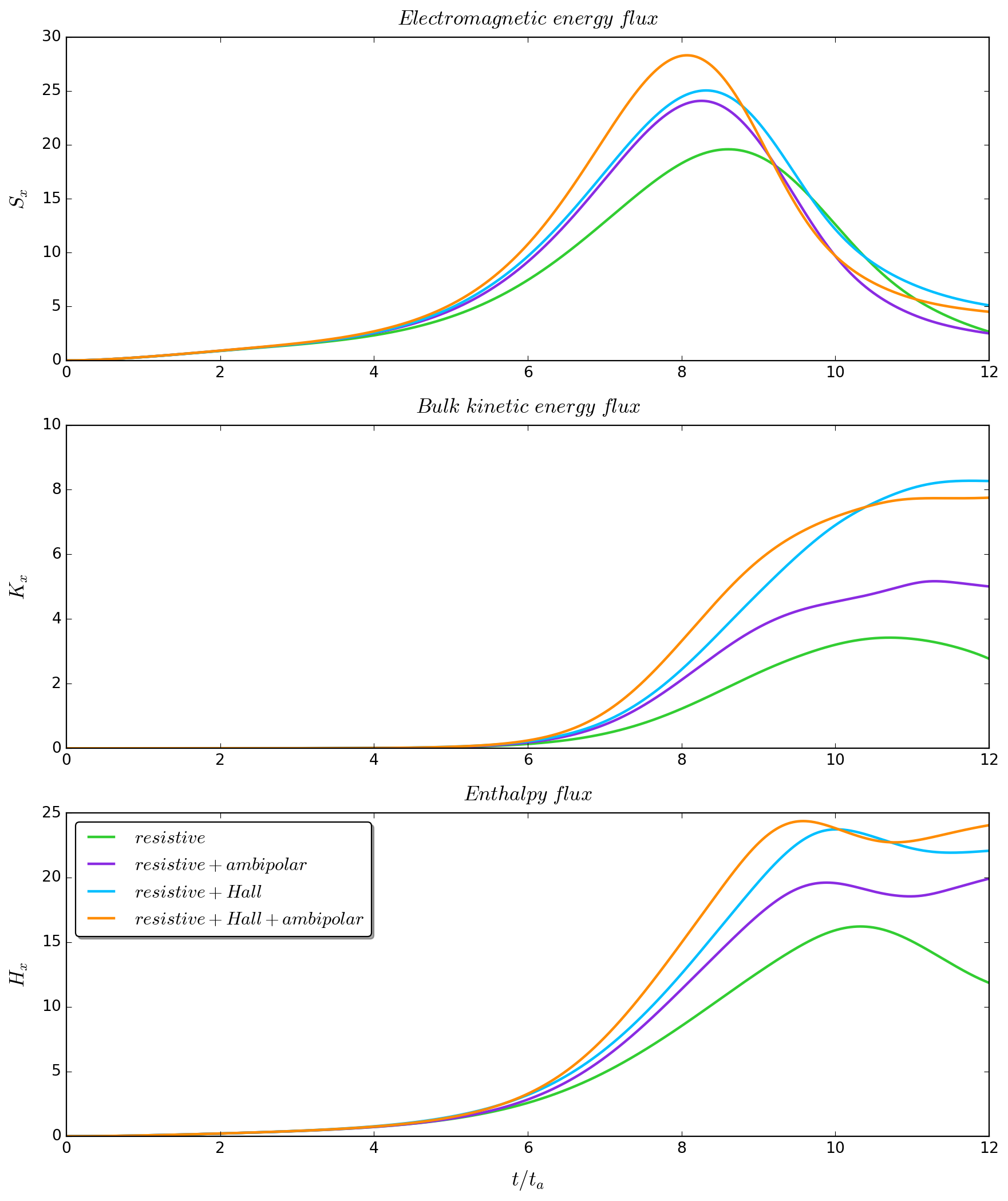}
\caption{Temporal evolution of the electromagnetic energy flux (first panel), bulk kinetic energy flux (second panel) and enthalpy flux (third panel).}
\label{fig_12}
\end{center}
\end{figure}

To fully understand both energy conversion and transport, it is necessary to consider not only the change in energy stored in a volume but also the amount of energy flowing through the surface surrounding it. In particular, we focus on a small region of the domain, including the diffusion region where reconnection occurs, and study the influence of the Hall effect and ambipolar diffusion on electromagnetic energy, bulk kinetic energy and enthalpy fluxes. The fluxes are related to the energy transport equation (\ref{eq.energy}) which, in addition to describing the conservation of energy, tells us how energy is transferred from one form to another. Specifically, the electromagnetic energy flux is given by the Poynting vector $\mathbf{S} = (\mathbf{E} \times \mathbf{B}) / \mu_{0}$, where the electric field can be obtained from Ohm's law (\ref{eq.Ohm}). The bulk kinetic energy flux is given by $\mathbf{K} = \rho u^{2} \mathbf{u} / 2$ and the enthalpy flux by $\mathbf{H} = (p+\rho e)\mathbf{u} = \gamma p \mathbf{u} / (\gamma - 1)$, which has a contribution from the internal energy density and another from the compressional work, $p \mathbf{u}$. These three quantities are integrated over a volume whose domain is given by $[8l_{a},16l_{a}] \times [-3.5l_{a},3.5l_{a}] \times [-0.1l_{a},0.1l_{a}]$. The results are shown in Figure \ref{fig_12}, where we have plotted these fluxes as a function of time for each case, i.e. resistive, resistive+Hall, resistive+ambipolar and resistive+Hall+ambipolar.

The results depicted in Figure \ref{fig_12} reveal that energy transport associated with reconnection primarily arises from electromagnetic and enthalpy fluxes, while kinetic energy flux plays a relatively minor role. Notably, the three panels in Figure \ref{fig_12} indicate that the rate of energy transfer is significantly higher in the resistive+Hall and resistive+Hall+ambipolar cases. This suggests that Hall and ambipolar effects may account for the rapid energy release during magnetic reconnection.

To quantitatively analyze this phenomenon, we measured the increases in $S_{x}$, $K_{x}$, and $H_{x}$ for the resistive+Hall, resistive+ambipolar, and resistive+Hall+ambipolar cases compared to the resistive case at time $t=8t_{a}$. For $S_{x}$, we observed a 29$\%$ increase in the ambipolar case, 34$\%$ in the Hall case, and 55$\%$ in the Hall+ambipolar case. Similarly, for $K_{x}$, the increases were 73$\%$, 99$\%$, and 171$\%$, respectively. Finally, for $H_{x}$, the increases were 33$\%$, 47$\%$, and 76$\%$, respectively. These measurements, conducted at time $t=8t_{a}$ and calculated relative to the resistive case, demonstrate that the resistive+Hall+ambipolar case exhibits the highest percentage increases, followed by the resistive+Hall case. Consequently, we can conclude that the Hall effect, particularly in combination with ambipolar diffusion, facilitates more efficient energy transport during reconnection.

\section{Conclusions}
\label{sec:conclusions}

Existing research has mainly focused on the Hall effect or ambipolar diffusion separately, without exploring the potential impact of their combined influence in a systematic way. In this work, we conducted a systematic study to analyze the effects of the Hall term and ambipolar diffusion on the formation and evolution of current sheets, the reconnection rate and the energy released during the magnetic reconnection process. Our findings underscore the importance of considering both effects in partially ionized plasma dynamics, where ambipolar diffusion cannot be neglected. We have observed significant enhancements in reconnection rates compared to simulations neglecting one or both processes. This integrated approach provides valuable insights into the acceleration of reconnection events in astrophysical plasmas, contributing to a more comprehensive understanding of magnetic reconnection speed-up mechanisms essential for the study of all astrophysical plasmas.

We solved the system of MHD equations with resistivity, Hall effect and ambipolar diffusion using MAGNUS. Since the code only solved the resistive MHD equations, we implemented new subroutines to calculate all terms related to the Hall effect and ambipolar diffusion in the numerical fluxes. The inclusion of these effects also required modifications to the time step. However, the presence of the Hall term in the code results in a significant reduction of the time step, which makes the execution time excessive, especially when a high spatial resolution is required.

Once the subroutines were implemented and modified, we used the initial data and boundary conditions proposed by the GEM magnetic reconnection challenge, a project to study the Hall effect in magnetic reconnection that has been used by many authors as a benchmark test. We performed four types of simulations, systematically turning on and off the Hall and ambipolar terms. Here, systematic means that the initial data, boundary conditions, computational domain, spatial resolution, and numerical methods were the same for all simulations, except for the Hall and ambipolar parameters, which changed depending on which term was turned on or off in each simulation. The four types of simulations were: resistive, resistive+Hall, resistive+ambipolar, and resistive+Hall+ambipolar.

First, we tested the Hall effect in MAGNUS by replicating the results of the GEM project. For the resistive and resistive+Hall simulations, we obtained a reconnected flux of 3.3 and 0.5, respectively, the same values as reported by \cite{GEM2001} in the GEM summary. Not only the values were the same, but also the qualitative behaviour of the curves. The same is true if we compare the results with those of other authors, such as \cite{Schmitz2006, Toth2008, Strumik2017}, where only slight differences are found. Finally, since we obtained the same results as in the test, MAGNUS shows fidelity in solving the MHD equations with the Hall term.

The ambipolar diffusion term was programmed similarly to the Hall effect. No benchmark test was performed this time, but we did investigate the effect of ambipolar diffusion on the GEM problem for the first time. This was done by systematically comparing simulations where only the ambipolar parameter $K_{A}$ was changed. In both the resistive+ambipolar and resistive+Hall+ambipolar simulations, the reconnected flux grows with the value of $K_{A}$. The higher value of $K_{A}$ was 0.01, the less ionized case compared to the other values used in this investigation. With $K_{A}=0.01$ we obtained a $75\%$ increase of the reconnected flux for the resistive+ambipolar simulation compared to the resistive one. For the resistive+Hall+ambipolar we got a $143\%$ increase in reconnected flux compared to the resistive+Hall simulation. However, after the increase, the flux of the latter remained quasi-stationary due to the formation of a plasmoid.

In general, the results of the GEM scenario show that the inclusion of the Hall effect produces a significant increase in the reconnected flux. Similarly, the inclusion of ambipolar diffusion produces significant increases. This suggests that ambipolar diffusion, like the Hall effect, could be one of the mechanisms for fast reconnection rates. %Moreover, ambipolar diffusion has not been studied in previous magnetotail investigations, and MMS spacecraft data from 2017 has revealed an ambipolar electric field transverse to the magnetic field in the magnetotail \cite{dubois2022}.

To study a model different from the GEM, which does not have a guide field, has uniform resistivity and has the current sheet formed since the beginning, we implemented a simple magnetic field configuration without perturbations, but with a guide-field and localized resistivity as initial data for the formation of a current sheet. In this model, the resistive+Hall+ambipolar case was the only one that reached a $0.1$ value in the reconnection rate, followed by the resistive+Hall case. These results show that the Hall effect is indeed the dominant phenomenon when it comes to increasing the reconnection rate and obtaining values close to $0.1$. However, ambipolar diffusion is also relevant, since it was the combination of the Hall effect and ambipolar diffusion that reached a maximum value of $0.1$. Moreover, simulations with ambipolar diffusion show a faster growth of the reconnection rate, which helps explain the rapid energy release that accelerates particles.

We also found that the conversion from magnetic to kinetic energy is more efficient in the presence of the Hall effect and its combination with ambipolar diffusion. This supports the fact that both mechanisms have a significant impact on particle acceleration processes. 

In addition to energy conversion, we also study its transfer through the fluxes present in the energy transport equation, and we found that energy transport is mainly due to electromagnetic and enthalpy fluxes. The kinetic energy flux is also present but is smaller than the others. For the three fluxes, the rate of energy transfer is higher in the resistive+Hall+ambipolar case, followed by resistive+Hall and resistive+ambipolar cases. Therefore, we conclude that both mechanisms provide more efficient energy transfer, especially when combined. Since the highest energy transfer rate is for the resistive+Hall+ambipolar case, we highlight the importance of ambipolar diffusion. 

Our results show that ambipolar diffusion causes magnetic energy to dissipate rapidly, facilitating the reconnection of magnetic field lines, and leading to the formation of thin current sheets. It is worth mentioning that the Hall effect enhances reconnection by altering the magnetic field topology and favouring the appearance of localized plasma current sheets, a natural outcome of the presence of a Burgers-like term. The above since the Hall term is proportional to the current, it is quadratic in $\mathbf{B}$, and hides a Burgers-like behaviour \citep{vainshtein2000rapid}). On the other hand, the ambipolar diffusion term, which is not quadratic but cubic in $\mathbf{B}$, is related to a velocity in the same direction as the Lorentz force. Therefore, its effect is to dissipate the currents perpendicular to B, acting to align magnetic and current fields \citep{Vigano2019}. Ambipolar effect plays a crucial role in the study of magnetic reconnection for several reasons. Mainly, it facilitates energy dissipation, which, together with the Hall effect, is essential for the optimal conversion of magnetic energy into thermal and kinetic plasma energy. The above is supported by our last results, where the rate of energy transfer shows a notable increase in the cases combining Hall and ambipolar effects.

Finally, it is essential to emphasize that although plasmas are often treated as single fluids, the distinction between ions, electrons, and neutral species becomes significant in diffusion regions where magnetic reconnection occurs. Therefore, future research should focus on a multi-species charged fluid model rather than a single-fluid approach. The different species interact through elastic and inelastic collisions. Elastic collisions involve the exchange of momentum and energy between different fluids, while inelastic collisions involve processes such as ionization and recombination. Adopting a multi-fluid approach allows for a more comprehensive study of regions within the flow sheet where turbulence is induced, such as the extremes where abrupt changes in velocity magnitude and direction occur. A recent study by \cite{Shibata2022} examined such regions in solar flares, where turbulence is also generated. A systematic study of turbulence excitation in the Earth's magnetotail using a multi-fluid framework would be very interesting and provide valuable insights. The above requires a model of the Earth's magnetotail current sheet without oversimplification.

\section*{Acknowledgements}

F.D.L-C was supported by the Vicerrectoría de Investigación y Extensión - Universidad Industrial de Santander, under Grant No. 3703.

%% The Appendices part is started with the command \appendix;
%% appendix sections are then done as normal sections
\appendix

\bibliographystyle{elsarticle-harv} 
%\bibliography{references}

\end{document}